\documentclass[acmsmall, screen]{acmart}
\acmJournal{TOSEM}

\usepackage{fontawesome}
\usepackage{float} 
\usepackage{pifont}
\usepackage{empheq}
\usepackage{graphicx}
\graphicspath{{./Figures/}} %
\usepackage[export]{adjustbox}
\usepackage{rotating}
\renewcommand{\arraystretch}{1.4}

\usepackage{color}

    {}   

\usepackage{color}
\usepackage{soul} %

\usepackage{enumitem}
\usepackage{colortbl}
\definecolor{gray50}{gray}{.5}
\definecolor{gray40}{gray}{.6}
\definecolor{gray30}{gray}{.7}
\definecolor{gray20}{gray}{.8}
\definecolor{gray15}{gray}{.85}
\definecolor{gray10}{gray}{.9}
\definecolor{gray05}{gray}{.95}

\usepackage{multirow}
\usepackage{tabularx}

\usepackage{hyperref}
\usepackage{xurl}

\usepackage{subcaption}

\usepackage{framed}
\usepackage{empheq}
\definecolor{mygray}{gray}{0.9}

\usepackage{amsmath}
\usepackage{amssymb}

\usepackage{rotating}
\usepackage{graphicx}
\usepackage{pdflscape}
\usepackage{longtable}
\usepackage{booktabs}
\usepackage{lscape}       %
\usepackage{algorithm}
\usepackage{algpseudocode}

\usepackage{adjustbox}

\usepackage{pgfplots}
\pgfplotsset{compat=1.18}
\usepackage{pgfplotstable}
\usepackage{adjustbox}

\definecolor{acm-blue}{HTML}{0F75BC}

\usepackage{tikz}
\newcommand{\sbar}[2]{%
  \begin{tikzpicture}[baseline]
    \pgfmathsetmacro{\maxwidth}{2.5} %
    \pgfmathsetmacro{\width}{\maxwidth * (#1 / #2)} %
    \fill[acm-blue] (0,0) rectangle (\width, 0.3); %
  \end{tikzpicture}%
}

\definecolor{steelblue}{RGB}{70,130,180}
\definecolor{darkgrey}{RGB}{169,169,169}

\newenvironment{boxC}{
    \MakeFramed{\hsize\linewidth\advance\hsize-\width\FrameRestore}\noindent%
}{%
    \endMakeFramed
}

\newcounter{keyTakeAwaysCounter}
\newenvironment{keyTakeAways}[1][Key Take Away]{
    \addtocounter{keyTakeAwaysCounter}{1}
    \begin{boxC}
    \faLightbulbO~\thekeyTakeAwaysCounter. \textbf{#1}: 
}{
    \end{boxC}
}

\newcounter{implicationsCounter}
\newenvironment{implications}[1][Implications]{
    \addtocounter{implicationsCounter}{1}
    \begin{boxC}
    \faBook~\theimplicationsCounter. \textbf{#1}: 
}{
    \end{boxC}
}

\begin{document}

\title{What Were You Thinking? An LLM-Driven Large-Scale Study of Refactoring Motivations in Open-Source Projects}
\author{Mikel Robredo}
\orcid{0009-0001-9870-1504}
\affiliation{
  \institution{University of Oulu}
  \city{Oulu}
  \country{Finland}
}
\email{mikel.robredo@oulu.fi}

\author{Matteo Esposito}
\orcid{0000-0002-8451-3668}
\affiliation{
  \institution{University of Oulu}
  \city{Oulu}
  \country{Finland}
}
\email{matteo.esposito@oulu.fi}

\author{Fabio Palomba}
\orcid{0000-0001-9337-5116}
\affiliation{%
  \institution{University of Salerno }
  \city{Salerno}
  \country{Italy}
}
\email{fpalomba@unisa.it}

\author{Rafael Pe\~naloza}
\orcid{0000-0002-2693-5790}
\affiliation{
  \institution{University of Milano-Bicocca}
  \city{Milano}
  \country{Italy}
}
\email{rafael.penaloza@unimib.it}

\author{Valentina Lenarduzzi}
\orcid{0000-0003-0511-5133}
\affiliation{
  \institution{University of Oulu}
  \city{Oulu}
  \country{Finland}
}

\begin{abstract}
\textbf{Context}. Code refactoring improves software quality without changing external behavior. Despite its advantages, its benefits are hindered by the considerable cost of time, resources, and continuous effort it demands.

\noindent\textbf{Aim}. Understanding why developers refactor, and which metrics capture these motivations, may support wider and more effective use of refactoring in practice.

\noindent\textbf{Method}. We performed a large-scale empirical study to analyze developers’ refactoring activity, leveraging Large Language Models (LLMs) to identify underlying motivations from version control data, comparing our findings with previous motivations reported in the literature.

\noindent\textbf{Results}. LLMs matched human judgment in 80\% of cases, but aligned with literature-based motivations in only 47\%. They enriched 22\% of motivations with more detailed rationale, often highlighting readability, clarity, and structural improvements. Most motivations were pragmatic, focused on simplification and maintainability. While metrics related to developer experience and code readability ranked highest, their correlation with motivation categories was weak.

\noindent\textbf{Conclusions}. We conclude that LLMs effectively capture surface-level motivations but struggle with architectural reasoning. Their value lies in providing localized explanations, which, when combined with software metrics, can form hybrid approaches. Such integration offers a promising path toward prioritizing refactoring more systematically and balancing short-term improvements with long-term architectural goals.

\end{abstract}


\maketitle

\section{Introduction}
\label{sec:Intro}
Maintaining codebases in the era of computing pervasiveness in daily tasks, from smart appliances to autonomous vehicles, is becoming increasingly challenging \citep{esposito2023uncovering}. With years of development and increasing reduction of time-to-releases, developers can make poor design choices to meet deadlines, leading to complex and hard-to-maintain software, thus increasing subsequent maintenance effort and costs for software firms \citep{lenarduzzi2021systematic}.
Code refactoring is one of the most well-known techniques to mitigate software complexity~\citep{fowler1999refactoring}. Refactoring aims to introduce code structure changes without altering external behavior, from code optimization to architecture and design patterns \citep{peruma2022refactor}. 

Developers restructure the code to improve its design, readability, and maintainability to ensure the long-term sustainability of software systems \citep{kim2012field, LACERDA2020110610,szoke2014bulk}. 
Thus, developers perceive it as a significant time and resource allocation cost due to its complexity and limitations \citep{kim2012field,avgeriou2020overview}. Therefore, with limited time-to-release windows, developers often limit resources spent on refactoring, hindering the quality and maintenance of the code \citep{avgeriou2020overview}. 
However, reducing software complexity with minor code refactoring can yield little to no improvement, leading to decreased code quality. In contrast, bulk and more comprehensive refactorings have a more profound influence on code quality \citep{LACERDA2020110610,szoke2014bulk}.

In an era in which developers benefit from the natural language processing capabilities LLMs bring when performing code refactoring~\citep{shirafuji2023refactoring, pomian2024assist, sergeyuk2024using, kim2025comparative}, understanding the best set of features that help provide the best refactoring recommendation can be challenging~\citep{peruma2022refactor}. Thus, our work focuses on identifying the motivations leading developers to perform code refactoring by leveraging LLMs, and similarly, compares the extracted motivations with those validated by the software engineering literature~\citep{silva2016we, pantiuchina2020developers}. Moreover, our study defines a catalogue with the most informative  software metrics (SMs) related to the identified refactoring motivations (RMs), thus aiding the development of refactoring recommendation systems tailored to existing needs and developer practices.

Our work is based on \citet{silva2016we} and extends \citet{pantiuchina2020developers}. More precisely, and comparing our work with the latter, we highlight the following notable differences concerning the study design and goals:
\begin{itemize}
    \item \textbf{Study design: Commit and Pull Requests (PR) Analysis.} While Pantiuchina et al.~\citep{pantiuchina2020developers} analyzed PR based on keyword lookup and the \textsc{RefactoringMiner} tool (RMT)~\citep{AlikhanifardTOSEM2024RefactoringMiner3} to detect refactoring-related commits, we propose a \textbf{novel approach based on LLMs} instead of keyword analysis. PRs represent developer interaction but cannot represent the full project change history \citep{yu2014reviewer,rahman2014insight}. PRs can be rejected; thus, it is also challenging to discern expert contributions among PRs accepted and not accepted ones~\citep{YU2016204}. Therefore, our work focuses on using all refactoring commits and not only on PRs.
    \item \textbf{Study design: Quality Metrics vs Product and Process Metrics.} Pantiuchina et al.~\citep{pantiuchina2020developers} focused on product-quality metrics such as  Object Oriented Programming (OOP) complexity, static analysis warnings, and developer-related metrics. We decided to focus on \textbf{established product and process metrics} as suggested by \citet{kamei2012large,rahm2013} as well as the ones introduced by Pantiuchina et al.~\citep{pantiuchina2020developers}.
    \item \textbf{Goals:} Pantiuchina et al.~\citep{pantiuchina2020developers} were interested in assessing the \textbf{correlation} between product quality and developer-related metrics and the occurrence of at least one refactoring in a specific PR. However, although PR is a richer container of developer thoughts and reflections, refactoring does not happen only in PR. Our study focuses on a broader set of metrics, analyzing qualitatively and quantitatively the \textbf{motivation} that pushes developers to refactor code, and thus \textbf{define a catalogue} of motivations supported by a large number of metrics, regardless of the connection of a specific PR.
\end{itemize}

To the best of our knowledge, no previous study identified the best performing SMs supporting refactoring recommendation by exploiting LLMs to analyze developers' commit messages and extract the code RMs. Our novel approach, which departs from conventional techniques, promises to deliver the first developer-based RMs to effectively target the correct amount of refactoring effort toward the right target. Our approach benefits practitioners, translating into a cost reduction, prioritizing refactoring efforts, and researchers paving the way for future works in developer-based software maintenance. 

We designed and conducted a large empirical study analysing \textbf{114 Java projects} hosted on GitHub.\footnote{\url{https://github.com}} Using RMT, we detected and analysed \textbf{13,725,139 refactoring activities} across a comprehensive list of \textbf{783,002 commit histories}. We then used advanced LLMs to interpret the motivations of developers expressed in version control history data, analysing their correlations with a robust set of established product and process metrics. This systematic approach enabled us to quantify the effectiveness of these metrics in capturing developer motivations and contributed to the development of a refined catalogue for developer-centric refactoring recommendation systems. Our study contributions are as follows:

\begin{itemize}
    \item We evaluate the alignment between LLM-generated refactoring motivations and motivations reported in Silva's work~\citep{silva2016we}, using expert judgment and statistical agreement analysis.
    \item We categorize and quantify how LLMs enrich or extend existing motivations with more detailed or context-aware rationale.
    \item We perform a large-scale open coding of LLM-derived refactoring motivations into 14 categories and analyze their distribution and dominant trends.
    \item We assess the information power and correlation of process and product SMs in explaining refactoring motivations using Random Forest, Extreme Gradient Boost, and statistical correlation tests.
    \item We discuss the implications for researchers and practitioners, outlining directions for LLM-guided refactoring recommendation systems.
\end{itemize}

The application of the undergone study revealed that LLMs agreed with human judgment in 80\% of the analyzed cases. However, only 47\% of the RMs identified by the LLMs fully aligned with those reported by the reference study~\citep{silva2016we}. Interestingly, in 22\% of the cases, LLMs extended known motivations with richer explanations, highlighting concerns like readability, testability, or naming clarity that were often implicit in prior work. A majority of motivations, over 55\%, centred around improving code clarity or reducing redundancy, revealing a predominantly pragmatic intent behind refactoring. While developer experience and readability-related metrics emerged as important factors in machine learning models, their direct correlation with motivation categories was statistically weak. Our study confirms the potential of LLMs to identify developer rationale in localized refactorings, while also exposing their limitations in capturing architectural intent, underscoring the need for hybrid systems that blend LLM reasoning with contextual project data.

\textbf{Paper Structure}. In Section~\ref{sec:CS}, we present the study design, while Section~\ref{sec:Results} presents the results and Section~\ref{sec:discussion} discusses them. Section~\ref{sec:Threat} focuses on threats to the validity of our study. Section~\ref{sec:relworks} discusses related work, and in Section~\ref{sec:Conclusions}, we draw the conclusions.

\section{Study Design}
\label{sec:CS}
\begin{figure}[t]
    \centering
    \includegraphics[width=\linewidth]{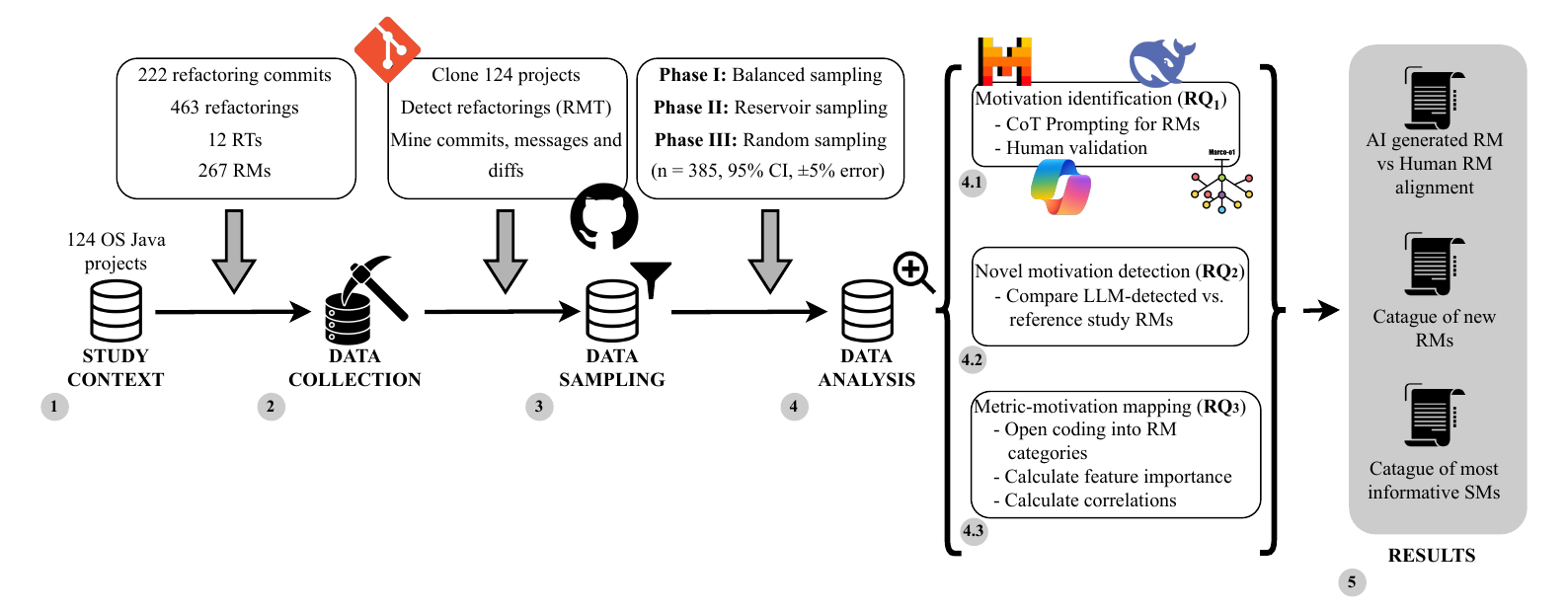}
    \caption{Workflow diagram of the study design.}
    \label{fig:study-design}
\end{figure}

This section outlines the empirical study, including the study goal and research questions, the study context, the data collection methodology, and the data analysis approach.
Our empirical study follows established guidelines defined by Wohlin et al. \citep{Wohlin2000}.  We publish the raw data in the replication package. Figure~\ref{fig:study-design} provides a graphical description of the study design undergone.

\subsection{Goal and Research Questions}
\label{sec:RQs}
The \emph{goal} of this study is to identify the motivations that drive developers to perform refactoring activities, with the \emph{purpose} of identifying the product and process metrics that can help understand the motivations pushing developers to perform refactoring. The \emph{perspective} is that of researchers and practitioners seeking additional support to guide developers in performing the refactoring. The \textit{context} is open-source Java projects.

Therefore, we derived the following research questions (RQs):

\begin{boxC}
\textbf{RQ$_1$.} Do motivations for refactoring in past studies align with those found in software project change histories?
\end{boxC}

Silva et al.~\citep{silva2016we} proposed a list of motivations for refactoring efforts provided by professional developers. Pantiuchina et al.~\citep{pantiuchina2020developers} built upon Silva et al.'s motivations, focusing on quality metrics and static analysis warnings, extracting information from PRs. 
However, PRs represent developer interaction but cannot represent the full project change history \citep{yu2014reviewer,rahman2014insight}. Moreover, PRs can be rejected; thus, it is also challenging to discern expert contributions among PRs accepted and not~\citep{YU2016204}. Hence, we investigate developer motivation, analyzing the single commits from the projects analyzed in the reference study~\citep{silva2016we} with a broader selection of product and process metrics. No prior studies considered employing the fully established product and process metrics~\citep{kamei2012large,rahm2013} with the refactoring history.
Furthermore, developers' commits, PRs, and code comments may not be able to capture all the motivations pushing developers to refactor the codebase; hence, we ask the following:

\begin{boxC}
\textbf{RQ$_2$.} Are there additional motivations driving the developers' willingness to perform refactoring?
\end{boxC}

Refactoring improves source code quality~\cite{peruma2022refactor}; yet, quality improvements in readability, performance, safety, and security are only part of the overall picture~\cite{palomba2017exploratory}. For instance, in an industrial context, it is essential to assess refactoring opportunities in situations with limited resources regarding work allocation and time \cite{szHoke2014case}. Silva et al.~\citep{silva2016we} succeeded in identifying 44 refactoring RMs at the time they ran their study. Therefore, we investigate whether external limitations or additional motivations influence refactoring efforts. 
Finally, extending Pantiuchina et al.,~\cite{pantiuchina2020developers} metric assessment, we ask the following:

\begin{boxC}
\textbf{RQ$_3$.} To what extent can product and process metrics reflect the motivations driving the developers' willingness to perform refactoring?
\end{boxC}

Based on \citet{pantiuchina2020developers}, a relationship exists between specific quality metrics and refactoring operations. However, their metric selection was limited to quality metrics and static analysis warnings. We extended their selection with established product and process metrics \citep{kamei2012large,rahm2013}, as well as a bigger body of data used in our study, given the time frame between Pantiuchina's work and the present study,  thus broadening the candidate for a plausible relationship to RMs. 
Therefore, we hypothesize that there must already be existing metrics in the commit history of software projects, highly correlated to the ground truth motivations and the hypothetically newly identified ones. The formulated hypotheses are defined as follows: 

\begin{itemize}
    \item \textbf{H}$_{0\mathcal{C}}$. There is no statistically significant correlation between the considered metrics and RMs.
    \item \textbf{H}$_{1\mathcal{C}}$. There is a statistically significant correlation between the considered metrics and RMs.
\end{itemize}

Consequently, considering \emph{H$_{1\mathcal{C}}$} as the accepted hypothesis and therefore the evidence of correlation, we expand our hypotheses as follows:

\begin{itemize}
    \item \textbf{H}$_{1\mathcal{C}.1}$. The identified correlation is negative.
    \item \textbf{H}$_{1\mathcal{C}.2}$. The identified correlation is positive.
\end{itemize}

In this sense, this question aims at identifying the metrics that can explain the existence of the stated motivations and further understand their specific implications. And therefore provide a list of metrics to monitor and guide developers for refactoring recommendations.

\subsection{Study Context}
\label{sec:Context}

In this study, we considered as the context dataset the one published by~\citet{silva2016we}, which consists of a catalogue of 44 distinct motivations for 12 well-known refactoring types (RTs). In their work, they initially selected the top 1000 Java repositories in GitHub,\footnote{https://github.com} ordered in terms of popularity, and further filtered them, removing the lower quartile based on the number of commits, thus obtaining a list of 748 repositories with consistent maintenance activities. Additionally, they filtered out from the datasets such projects which did not show development activity while conducting their study, thus reducing the number of projects to 471. 
The final set of considered projects comprised 124 repositories, which featured at least one refactoring during the study period, and answers provided by developers regarding the specific refactoring activity to the author's questions on the refactoring motivation.
The final set of projects comprised 222 commits, for which the authors provided answers for a total of 267 RMs, and 44 unique motivations. Within this set of commits, 463 refactorings from 12 RTs were detected using the RMT~\citep{tsantalis2013multidimensional}. 
Therefore, to validate the motivations provided by developers, our study considers the final set of 124 GitHub-hosted Java projects described in the adopted dataset.\footnote{http://aserg-ufmg.github.io/why-we-refactor} 

In Table~\ref{tab:study-context}, we present the descriptive statistics of the projects that completed all the stages in the data collection (see Section~\ref{sec:DataCollection}). We can observe that the population of repositories presents an average maturity of more than 13 years, with some projects indicating a maximum age of more than 23 years. Regarding the popularity of the studied projects, measured in terms of GitHub stars, we also observed a highly skewed distribution, with the most stargazed project holding 78,022 stars, while the mean remained at 7,212.38 and the median at 3,603. We also revealed extreme dispersion among projects based on characteristics such as the number of commits, which presented an average number of almost 21,000 commits, but with some projects containing as many as 401,000 commits. Similarly, the number of developers involved in a project varied from 11 to over 2,300, and the number of programming languages ranged from 1 to 22, thus depicting diverse levels of team size and project complexity. To compute the number of programming languages, we only considered programming languages following the criteria adopted from the TIOBE index.\footnote{\url{https://www.tiobe.com/tiobe-index/programminglanguages_definition/\#instances}} According to these criteria, a programming language should have an entry in Wikipedia that defines it as a programming language, and further, it should have at least 5,000 results for Google search \texttt{<language> programming}. Furthermore, this index only considers programming languages that are \textit{Turing complete}~\citep{stuart2013understanding}. The TIOBE index contains 358 programming languages, which we provided in the replication package of this study (see Data Availability Statement).

Moreover, we computed the project's size based on effective lines of code (ELOC), which resulted in a highly skewed distribution of the projects (kurtosis = 36.26). Peak-sized projects presented almost 11 million ELOC, while the average size remained in more than 570,000 ELOC, which already denoted an average noticeable size of the considered projects. To compute ELOC,  we quantified the total number by only considering the programming languages included in the TIOBE index. Altogether, we consider the study context for this study to be representative of an extended variety of Java projects in terms of the presented software attributes.

\begin{table}
\centering
\color{black}
\caption{Summary descriptive statistics of the final set of projects considered in the study}
\label{tab:study-context}
\resizebox{\linewidth}{!}{%
\begin{tabular}{@{}lrrrrrrrr@{}}
\toprule
\multicolumn{1}{c}{} & \multicolumn{1}{c}{\textbf{Mean}} & \multicolumn{1}{c}{\textbf{Standard deviation}} & \multicolumn{1}{c}{\textbf{Median}} & \multicolumn{1}{c}{\textbf{Min}} & \multicolumn{1}{c}{\textbf{Max}} & \multicolumn{1}{c}{\textbf{Skew}} & \multicolumn{1}{c}{\textbf{Kurtosis}} & \multicolumn{1}{c@{}}{\textbf{Standard error}} \\ \midrule
\textbf{Age} & 13.67 & 3.99 & 14.13 & 2.86 & 23.81 & -0.36 & 0.18 & 0.38 \\
\textbf{Size*} & 572,311.79 & 1,367,904.93 & 174,573 & 4,221 & 10,745,685 & 5.73 & 36.26 & 129,835.79 \\
\textbf{Stars} & 7,212.38 & 11,908.57 & 3,603 & 351 & 78,022 & 4.24 & 20.23 & 1,120.26 \\
\textbf{Open issues} & 398.24 & 655.53 & 177 & 0 & 4,278 & 3.51 & 14.52 & 62.79 \\
\textbf{Closed issues} & 4,064.06 & 5,871 & 2,445 & 55 & 38,493 & 3.94 & 18.60 & 562.34 \\
\textbf{Overall issues} & 4,462.29 & 6,253.89 & 2,912 & 60 & 40,400 & 3.82 & 17.76 & 599.01 \\
\textbf{Commits} & 21,231.53 & 47,849.40 & 7,818 & 344 & 401,429 & 5.73 & 38.41 & 4,501.29 \\
\textbf{Developers} & 347.50 & 451.02 & 195 & 11 & 2,305 & 2.53 & 6.49 & 42.43 \\
\textbf{Languages} & 4.42 & 3.54 & 4 & 1 & 22 & 2.37 & 7.70 & 0.33 \\
\textbf{Refactoring Commits} & 92.35 & 10.91 & 96 & 47 & 103 & -1.93 & 3.83 & 1.03 \\
\textbf{Refactoring types} & 6,663.39 & 18,650.19 & 2,389 & 77 & 186,854 & 8.28 & 75.98 & 1,754.46 \\
\textbf{Refactorings} & 117,424.62 & 522,169.20 & 32,437 & 385 & 5,509,181 & 9.89 & 99.42 & 49,121.55 \\
\bottomrule
\end{tabular}%
}
\end{table}

\subsection{Data Collection}
\label{sec:DataCollection}

\begin{figure*}[t]
    \centering
    \includegraphics[width=\textwidth]{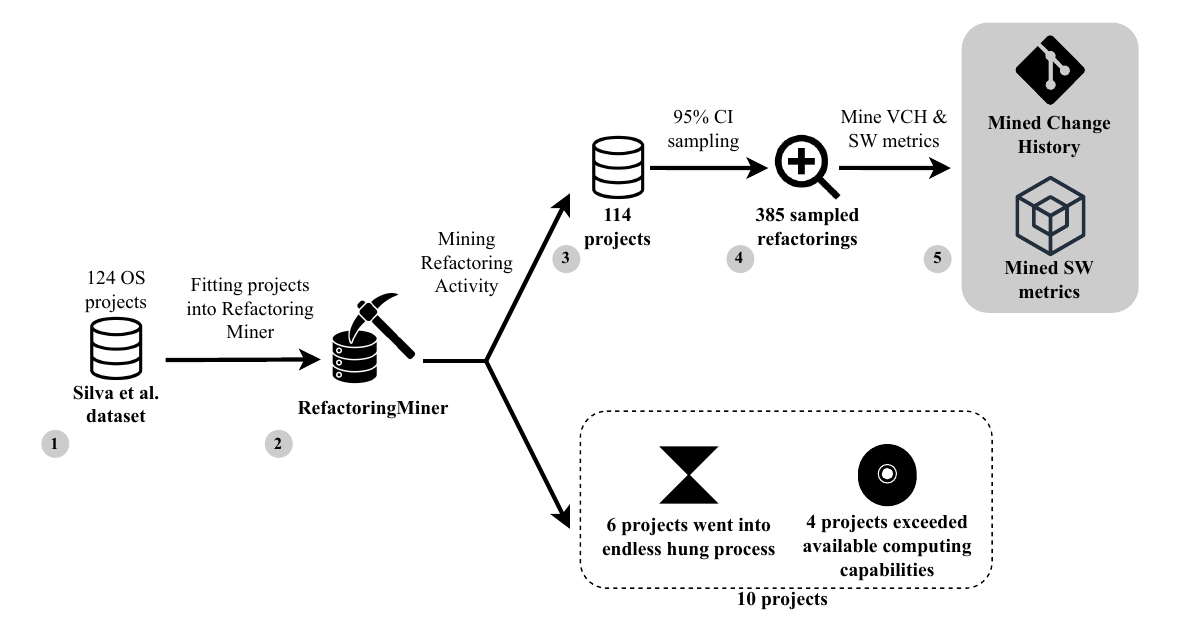}
    \caption{Data collection process diagram.}
    \label{fig:data_collection_diagram}
\end{figure*}

This section outlines our multi-stage data collection process, depicted in Figure~\ref{fig:data_collection_diagram}, and designed to support our RQs. The process includes: (i) mining refactoring data, (ii) creating a statistically significant sample for the analysis, (iii) mining project change history, and (iv) computing SMs listed in Table~\ref{tab:jit-metrics}. To aid the narrative of our study design, we have included further details on the data collection process in Appendix~\ref{sec:AppendixA}.

\subsubsection{Collecting Refactoring Data}
\label{sec:collecting-refactoring-data}

To ground our study in real-world developer activity, we mined the refactoring data from 124 GitHub projects, replicating and extending the dataset from the reference study~\citep{silva2016we}. Since their publication, these projects experienced an average increase of over 10,600 commits; e.g., \textit{JetBrains/intellij-community} increased by 273,044 commits. We provide a table presenting the differences between the number of commits in the studied projects when the reference study was published and for the present study in the replication package (see Data Availability Statement).

We adopted the \textsc{RefactoringMiner} tool (RMT)~\citep{AlikhanifardTOSEM2024RefactoringMiner3} to mine the refactoring activity from the entire version-control history of the cloned repositories. Table~\ref{tab:refactorings} shows the categorization of the 103 refactorings of our study according to their types as defined by Fowler~\citep{fowler1999refactoring}. 

Due to resource limitations and persistent errors with RMT, we excluded 10 projects from the study. However, since we could still collect the SMs from affected refactoring commits, we included the affected commits and their ground-truth RMs. We acknowledge the impact of this decision as a threat to validity in Section~\ref{sec:Threat}. We extend the description of the data collection process and issues in Appendix~\ref{sec:AppendixA_1}.

\begin{table}
\centering
\footnotesize
\caption{Refactorings Types detectable by \textsc{RefactoringMiner 3.0}  and newer versions~\citep{AlikhanifardTOSEM2024RefactoringMiner3}.}
\label{tab:refactorings}
\begin{tabular}{m{3.5cm}|m{3.5cm}|m{5cm}}
\hline 
\textbf{Group} &
\textbf{Description} &
\textbf{Refactorings} \\ \hline 
    \textbf{Composing Methods} &
    Re-organize how methods are composed, such as streamlining their logic or removing unneeded parts. &
    \textbf{Extract}~/~\textbf{Inline}~/~Merge~/~Split \textbf{Method}, Extract~/~Inline~/~Split~/~Merge~/~Rename Variable, Change Variable Type, Move Code (between methods), Merge Catch / Conditional, Split Conditional.
    \\ \hline
    \textbf{Moving Features between Objects} &
    Re-organize the distribution of functionalities and data among classes. &
    Extract~/~\textbf{Move}~/~Rename \textbf{Class}, \textbf{Move Method}, \textbf{Move Attribute}, Localize~/~Reorder Parameter, Replace Attribute~/~Attribute with Variable.
    \\ \hline
    \textbf{Manage Objects Modifiers} &
    Re-assign the modifiers from different parts of the code. & Change Attribute Access~/~Class Access Modifier, Change Type Declaration Kind, Add Method~/~Attribute~/~Variable~/~Parameter~/~Class Modifier, Remove Method~/~Attribute~/~Variable~/~Parameter~/~Class Modifier.
    \\\hline
    \textbf{Organizing Data} &
    Re-organize the way data is managed inside a class. &
    Extract~/~Split~/~Merge~/~Replace~/~Rename~/~Inline~/~Encapsulate~/~Parameterize Attribute, Change Attribute Type, Replace Variable With Attribute.
    \\\hline
    \textbf{Simplifying Method Calls} &
    Simplify class interactions by making the methods easier to call and understand. &
    Split~/~Merge~/~Add~/~Remove~/~Reorder~/~Rename Parameter, Parameterize Variable, Change Parameter Type, Change Method Access Modifier, Change Return Type, Rename Method.
    \\\hline
    \textbf{Dealing with Generalization} &
    Moving functionalities along class inheritance hierarchy. &
    \textbf{Extract Superclass}, Extract Subclass, \textbf{Extract Interface}, \textbf{Pull Up}~/~\textbf{Push Down} \textbf{Attribute}, \textbf{Pull Up}~/~\textbf{Push Down} \textbf{Method}, Split / Merge Class.
    \\\hline
    \textbf{Object Replacement} &
   Replace source code objects with alternatives. & Replace Loop With Pipeline / Anonymous With Lambda / Pipeline With Loop / Anonymous With Class / Generic With Diamond / Conditional With Ternary.
    \\\hline
    \textbf{Package Management} &
    Re-organize how packages are composed. & \textbf{Rename}~/~Move~/~Split~/~Merge \textbf{Package}.
    \\\hline
    \textbf{Test Specific} &
    Handle test specific scenarios. & Parameterize Test, Assert Throws~/~Timeout
    \\\hline
    \textbf{Others} &
    Other composite refactorings are detected by \textsc{RefactoringMiner}. &
    Move And Rename Attribute, Move And Inline Method, Move And Rename Class, Move And Rename Method, Extract And Move Method, Add~/~Modify~/~Remove (Class~/~Attribute~/~Method~/~Parameter~/~Variable) Annotation, Add~/~Change~/~Remove Thrown Exception Type, Change Package, Move Source Folder, Try With Resources, Invert Condition, Collapse Hierarchy.
    \\ \hline
    \multicolumn{3}{l}{(\textbf{Bold RTs}: RTs included in the reference study.)}
\end{tabular}%
\end{table}

In total, we mined 114 projects, detecting 13,725,139 refactorings across 783,002 commits. Commits without refactorings were discarded. Abbreviations for refactoring types (RTs) used throughout the article are listed in Appendix Table~\ref{tab:Abbreviations}. Refactoring category frequencies are shown in Table~\ref{tab:refactoring-counts}.

\begin{table}
\begingroup
\color{black}
\centering
\footnotesize
\caption{Mined frequencies for each RT detected by \textsc{RefactoringMiner} (\#: Instances).}
\label{tab:refactoring-counts}

\begin{tabular}{lr|lr|lr|lr|lr|lr}
\hline
\textbf{RT} & \textbf{\#} & \textbf{RT} & \textbf{\#} & \textbf{RT} & \textbf{\#} & \textbf{RT} & \textbf{\#} & \textbf{RT} & \textbf{\#} & \textbf{RT} & \textbf{\#} \\ \hline
RM    & 1,997,099 & MM    & 198,667 & RC    & 76,450 & RAA   & 35,118 & TWR   & 8,283  & SM    & 1,957 \\
MMA   & 1,663,555 & RCA   & 196,080 & AAA   & 76,300 & EC    & 34,097 & EI    & 8,123  & MerC  & 1,682 \\
CVT   & 694,433   & APA   & 182,867 & ATET  & 73,858 & IC    & 32,880 & CTDK  & 8,117  & MCat  & 1,677 \\
AMA   & 649,691   & RTET  & 181,235 & RPM   & 69,931 & CTET  & 28,877 & AVA   & 7,783  & CH    & 1,441 \\
CPT   & 605,118   & RA    & 171,885 & CCAM  & 62,162 & RAWV  & 28,764 & RParam& 7,739  & RPWL  & 1,427 \\
AP    & 504,837   & IV    & 165,687 & RAM   & 60,016 & RCM   & 27,218 & RAWC  & 5,173  & SClass& 1,016 \\
CRT   & 438,291   & ACA   & 144,488 & IM    & 59,572 & LP    & 22,840 & MA    & 4,932  & MerM  & 945 \\
EV    & 360,499   & CAAM  & 143,428 & PV    & 55,221 & MCode & 22,184 & AT    & 4,766  & SP    & 877 \\
MCA   & 342,591   & AVM   & 133,253 & RMM   & 55,131 & EA    & 21,651 & IA    & 4,752  & MPack & 705 \\
CMAM  & 330,173   & AAM   & 125,500 & ACM   & 49,746 & MCon  & 21,021 & SP    & 4,372  & SV    & 688 \\
EM    & 324,173   & RAWL  & 123,551 & EnA   & 47,433 & SC    & 18,579 & ESub  & 4,187  & RA    & 615 \\
RenP  & 297,012   & RVM   & 117,181 & MARM  & 46,839 & MSF   & 18,562 & MARA  & 4,124  & MVA   & 305 \\
RV    & 292,827   & RGWD  & 109,725 & RVWA  & 46,643 & Sup   & 16,199 & MP    & 3,828  & PT    & 73 \\
CAT   & 282,403   & PUM   & 102,417 & RPA   & 43,379 & PDA   & 15,160 & RVA   & 3,745  &  -     & -     \\
MARM  & 281,148   & APM   & 99,209  & PUA   & 42,097 & PA    & 11,683 & MV    & 3,418  &  -     & -     \\
RP    & 261,442   & MA    & 97,947  & PDM   & 39,711 & RCWT  & 10,976 & RLWP  & 3,326  &  -     & -     \\
MovC  & 251,219   & EAMM  & 88,375  & MAIM  & 38,629 & MParam& 10,815 & RPack & 3,088  &  -     & -     \\
RMA   & 212,805   & AMM   & 82,472  & MAA   & 36,976 & MPA   & 9,352  & SA    & 2,622  &  -     & -     \\
\hline
\end{tabular}

\endgroup
\end{table}

\subsubsection{Selecting a Statistically Significant Sample for the Analysis.}

Analyzing more than 13 million refactorings was computationally infeasible due to the cost of SM extraction (e.g., ADEV, MINOR) (see Section~\ref{sec:software-metrics}) and prompt-based LLM analysis (see Section~\ref{sec:DataAnalysis}). 

\begin{figure*}[t]
    \centering
    \includegraphics[width=\textwidth]{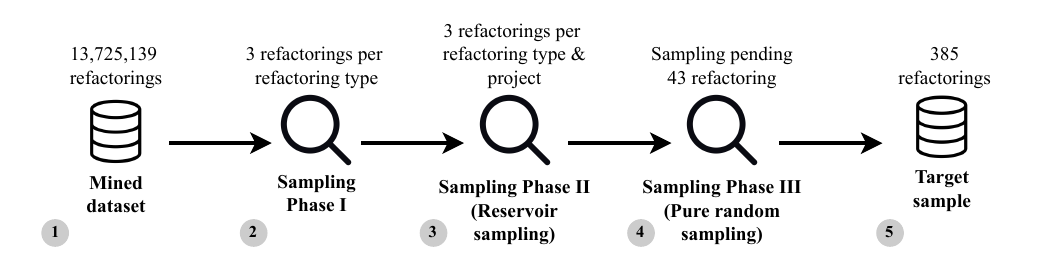}
    \caption{Diagram of the adopted sampling strategy.}
    \label{fig:sampling_process_diagram}
\end{figure*}

To ensure tractability and rigor, we selected a sample of 385 observations, yielding a 95\% confidence level and 5\% margin of error in the obtained results~\citep{billingsley2013convergence,sandelowski1995sample}. For that, we followed a similar sampling approach to that of already implemented sampling techniques within the field of Software Engineering~\citep{lenarduzzi2023critical} (see Figure~\ref{fig:sampling_process_diagram}). We applied a greedy sampling strategy over three phases:

\begin{itemize}
    \item \textbf{Phase 1}: Ensure each project and RT had a minimum of 3 sampled refactorings.
    \item \textbf{Phase 2}: Use reservoir sampling~\citep{vitter1985random} to fill gaps in underrepresented projects.
    \item \textbf{Phase 3}: Apply random sampling to complete the remaining unfilled entries.
\end{itemize}

We expand on the description of the performed sampling strategy in Appendix~\ref{sec:AppendixA_2}.

\subsubsection{Mining the Change History}

For each sampled commit, we used the PyDriller\footnote{\url{https://pydriller.readthedocs.io/en/latest/}} Python library to extract the commit message and the corresponding file-level code diff where the refactoring was applied. PyDriller also supports basic structural analysis using Lizard,\footnote{\url{https://github.com/terryyin/lizard}} enabling us to capture class- and method-level metrics across languages.

\subsubsection{Computing the SMs}
\label{sec:software-metrics}

To answer \textbf{$RQ_3$}, we computed both product and process metrics~\citep{kamei2012large,rahm2013} leveraging different tools. We used:

\begin{itemize}
    \item \textbf{PyDriller} for SMs requiring the cumulative inspection of the version-control history between commits. 
    \item \textbf{CK}~\citep{aniche_ck} for product metrics~\citep{pantiuchina2020developers} (e.g., WMC, RFC), invoked via CLI over target Java files.
    \item \textbf{CoRed},\footnote{\url{https://github.com/grosa1/CoRed/tree/master}} a Python wrapper of the official implementation of the metric described by~\citet{scalabrino2018comprehensive} to compute ComRead, which provides a comprehensive readability score.
\end{itemize}

Table~\ref{tab:jit-metrics} presents and describes all the metrics considered in this study. %

\begin{table}
\centering
\caption{Product and process metrics adopted in this study.}
\label{tab:jit-metrics}

\tiny
\begin{tabular}{@{}l|l|p{11cm}@{}}
\toprule
& \textbf{Metric} & \textbf{Description} \\ 
\midrule
\multirow{15}{*}{\rotatebox[origin=c]{90}{\cite{rahm2013}}} 
& COMM     & The cumulative number of commits in a given file up to the considered commit.                                                  \\
& ADEV     & The cumulative number of active developers who modified a given file up to the considered commit.                              \\
& DDEV     & The cumulative number of distinct developers contributed to a given file up to the considered commit.                          \\
& ADD      & The normalized number of lines added to a given file in the considered commit.                                                 \\
& DELE      & The normalized number of lines removed from a given file in the considered commit.                                             \\
& OWN      & Measures the percentage of the lines authored by the highest contributor of a file.                                                            \\
& MINOR    & The number of contributors who contributed less than 5\% of a given file up to the considered commit.                          \\
& SCTR     & The number of packages modified by the committer in the considered commit.                                                     \\
& NADEV    & The number of active developers who changed any of the files involved in the commits where the given file has been modified.   \\
& NDDEV    & The number of distinct developers who changed any of the files involved in the commits where the given file has been modified. \\
& NCOMM    & The number of commits where the given file has been involved.                                                                       \\
& NSCTR    & The number of different packages touched by the developer in commits where the file has been modified.                         \\
& OEXP     & The percentage of code lines authored by a given developer in the project.                                               \\
& EXP      & The mean of the experience of all developers across the project.                                                         \\
\midrule
\multirow{14}{*}{\rotatebox[origin=c]{90}{\cite{kamei2012large}}} 
& ND       & The number of directories involved in a commit.                                                                                \\                                       
& NS       & Number of modified subsystems (packages).                      \\
& NF       & Number of modified files. \\
& ENTROPY  & The distribution of the modified code across each given file in the considered commit.                                         \\
& LA       & Ten lines added to the given file in the considered commit (absolute number of the ADD metric).                      \\
& LD       & The number of lines removed from the given file in the considered commit (absolute number of the DEL metric).                  \\
& LT       & The number of lines of code in the given file in the considered commit before the change.                                      \\
& FIX      & Whether or not the change is a defect fix. \\
& NDEV     & The number of developers that changed the modified files. \\
& AGE      & The average period between the last and the current change.                                                                 \\
& NUC      & The number of times the file has been modified up to considered commit.                                                  \\
& CEXP     & The number of commits performed on the given file by the committer up to the considered commit.                                \\
& REXP     & The number of commits performed on the given file by the committer in the last month.                                          \\
& SEXP     & The number of commits a given developer performs in the considered package containing the given file.                   \\
\midrule
\multirow{19}{*}{\rotatebox[origin=c]{90}{\cite{pantiuchina2020developers}}} 
 & CBO & Coupling Between Object classes: measures the dependencies a class has. \\
 &         WMC & Weighted Methods per Class: sums the cyclomatic complexity of the methods in a class. \\
  &        RFC & Response For a Class: the number of methods in a class plus the number of remote methods that are called recursively [citeme]. \\
 &         ELOC & Effective Lines Of Code: the lines of code excluding blank lines and comments. \\
 &         NOM & Number Of Methods in a class. \\
 &         NOPM & Number Of Public Methods in a class. \\
 &         DIT & Depth of Inheritance Tree: the length of the path from a class to its farthest ancestor.  \\
 &         NOC  & Number Of Children (direct subclasses) of a   class. \\
 &         NOF & Number Of Fields declared in a class. \\
 &         NOSF & Number Of Static Fields declared in a class. \\
 &         NOPF & Number Of Public Fields declared in a class. \\
 &         NOSM & Number Of Static Methods in a class. \\
 &         NOSI & Number Of Static Invocations of a class. \\
 &         HsLCOM & Henderson-Sellers revised Lack of Cohesion   Of Methods (LCOM): a class cohesion metric based on sharing local instance variables by the class methods.\\ %
 & ComRead & Comprehensive readability model: combines structural, visual (e.g., alignment),   and textual features (e.g., comments readability).  \\
        
\bottomrule
\end{tabular}%

\end{table}

\subsection{Data Analysis}
\label{sec:DataAnalysis}

This section describes the data analysis techniques and methods used for answering our RQs, including the identification of RMs for each of the sampled refactoring observations (RQ$_1$), the alignment and identification of additional RMs compared to the reference study (RQ$_2$), as well as the evaluation of the SMs that best inform about the identified RMs (RQ$_3$). Figure~\ref{fig:data-analysis-diagram} presents the workflow diagram of the data analysis.

\subsubsection{LLM Roles, Human Validation and Prompting Strategy.}
\label{sec:LLMs}

Since the general availability of LLMs, our research community has investigated the use of LLMs in SE tasks, benefiting both practitioners and researchers \cite{esposito_generative_2025}. Among such tasks, classification and rating are the most time-consuming and error-prone tasks that humans perform during research activities \cite{esposito_large_2024, esposito_call_2025}. Therefore, and since we performed LLM prompting in different stages of the data analysis to answer our RQs, we evaluated the proficiency, accuracy, and agreement of four LLMs assigned to extract the motivation behind refactoring operations. The following paragraphs describe the roles of the LLMs adopted in the data analysis design, the human validation employed in each of the response collections obtained from the LLMs, as well as the prompting strategy adopted to guide the LLMs in performing the tasked classifications. 

\begin{figure}[t]
    \centering
    \includegraphics[width=0.8\linewidth]{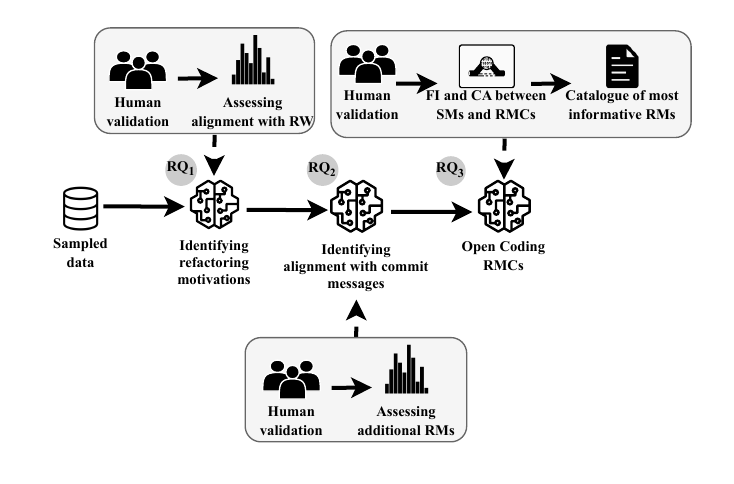}
    \caption{Data analysis process diagram.}
    \label{fig:data-analysis-diagram}
\end{figure}

\paragraph{\textbf{LLM Roles.}} We employed four distilled LLMs for classification and validation tasks:

\begin{itemize}
    \item \textbf{Large Reasoning Model (LRM):} \underline{Marco-o1} acted as the primary assistant for identifying RMs. Its output formed the basis for further validation and was referred to as the LRM response throughout the study.

    \item \textbf{First Validation Model (V1):} \underline{Mistral NeMo} was assigned as the first validation assistant. It received the same input as the LRM and was additionally asked to evaluate the LRM's reasoning and output. Its role was to independently assess the correctness of the LRM’s decision.

    \item \textbf{Second Validation Model (V2):} \underline{DeepSeek R1} acted as the second validation assistant, mirroring the role of V1. It provided a parallel evaluation of the LRM’s output to detect agreement or disagreement with V1’s assessment.

    \item \textbf{Third Validation Model (V3):} \underline{Microsoft Phi-4} was designated as the final arbiter in cases where V1 and V2 disagreed. In addition to the standard input, it received the assessments from both V1 and V2, as well as the original LRM output, to deliver a final decision in conflicting cases.
\end{itemize}
Each selected model was assigned a specific role based on its function in the data analysis pipeline (see Figure~\ref{fig:data-analysis-diagram}). We provide further technical details on the adopted LLMs, as well as model configuration in Appendix~\ref{sec:AppendixB_1}. 

\paragraph{\textbf{Human Validation.}} Since no previous study reported findings on the accuracy of LLMs for the task at hand, we designed a validation strategy involving three human experts replicating the LLM validator's roles. The goal was to assess the quality of the model-generated motivations and identify which models consistently produced reasonable outputs.
Our validation followed a three-step protocol: 
\begin{itemize}
    \item One expert independently reviewed the same input provided to the LRM and the three validation models (V1–V3), and manually evaluated the correctness of each model's motivation. For each case, and after making a manual decision over the refactoring case, the expert indicated whether they \texttt{agreed} or \texttt{disagreed} with the LRM’s motivation, noted the majority decision among the validation models, and identified the models they considered correct.
    \item  A second expert repeated the same evaluation independently and documented their level of agreement with the first expert’s judgments. 
    \item  In cases of disagreement between the first two reviewers, a third expert was brought in to assess the same outputs independently. Final decisions were made through majority voting among the three validators.   
\end{itemize}

Since LLM outputs are non-deterministic, and to assess the reliability of the results, we performed this validation strategy in each of the data analysis stages, which involved LLM outputs (see Figure~\ref{fig:data-analysis-diagram}). 

\paragraph{\textbf{Prompting Strategy.}} 

To guide LLMs in classifying RMs, we adopted chat-based in-context learning, which has been shown to perform comparably to fine-tuning~\citep{esposito_beyond_2024}. Each prompt consisted of two components: a \textbf{system message} defining the model’s role and expected output, and a \textbf{user message} providing the input context.

We employed prompt engineering techniques to guide LLMs in classifying RMs. According to the state-of-the-art, in-context learning through chat-based prompting provides similar or better results than the more computationally expensive fine-tuning process \cite{esposito_beyond_2024}. During the in-context learning phase, the system message established the assistant’s role and specified the expected output format, and the user message provided the contextual input.

The user message included details a human developer would typically see: the \texttt{RT}, a \texttt{description} (from RMT), the \texttt{commit message}, and the relevant \texttt{code diff}. Information on refactoring activity and local changes collected from the version-control history of a project provides the LLMs with localized context on the performed refactorings. However, this might lack more holistic details, such as long-term project goals or further architectural design information. We acknowledge this gap, and acknowledge it as a threat to validity in Section~\ref{sec:Threat}.

To improve the attention and consistency of the LLM, we adopted the \textit{Chain of Thought} (CoT) prompting~\citep{wei2022chain}, employing zero-shot learning~\citep{pourpanah2022review, kojima2022large}. We systematically refined the employed message prompts during the prompt engineering stage for each of the defined RQs, and made them accessible in our replication package (see Data Availability Statement). Similarly, we expand how we used the CoT prompting strategy, as well as the employed JSON prompt format in Appendix~\ref{sec:AppendixB_2}.

\subsubsection{Alignment and Extension of RMs with the Reference Work (RQ$_1$-RQ$_2$)} 
\label{sec:RQ1-RQ2-data-analysis}

To answer RQ$_1$, we prompted the LLMs with each of the 385 sampled observations (see Section~\ref{sec:DataCollection}), including the RT, RMT-generated description, commit message, and code diff. To evaluate the accuracy of the resulting RMs identified through the LLM validation, we compared them with the ground truth RM from the reference study.

This comparison involved checking for logical consistency, identifying discrepancies, and determining whether the identified RMs extended or complemented the ground truth~\citep{silva2016we}. It is important to note that while our validated dataset included all RTs mineable by the latest RMT version, the ground truth only covered those listed in Table~\ref{tab:refactoring-motivations-related}. Table~\ref{tab:Abbreviations} provides the names and abbreviations of such RTs.

Therefore, the validation process was limited to RTs analyzed in~\citet{silva2016we}. We paired the sampled RTs and their identified RMs with the ground truth RM counterparts, resulting in a total of 758 RM comparison pairs. From this set, we performed human validation on a statistically significant sample of 198 pairs, with a 95\% confidence level and 5\% margin of error~\citep{sandelowski1995sample} (see Section~\ref{sec:LLMs}).

While the goal of $RQ_1$ consisted of the alignment between the ground-truth RMs and those now identified, we needed to validate such results through manual validation, and further, assess the reliability of the performed validation. Therefore, and to assess the reliability of the validation, we measured the inter-rater agreement (IRA) between the two raters (human and LLM) using \textbf{Cohen’s Kappa} ($\kappa$)~\citep{cohen1960coefficient}, which is well-suited for two-rater scenarios. Unlike Fleiss’ Kappa~\citep{fleiss1971measuring}, which is designed for multiple raters, Cohen’s $\kappa$ allows us to measure the agreement between a single LLM and a human validator. We interpreted the results using Landis and Koch’s guidelines~\citep{landis1977measurement}, where $\kappa$ values range from less than 0 (no agreement) to 0.81–1 (almost perfect agreement).

Moreover, we conjectured the following hypotheses on the symmetry of the agreement:

\begin{itemize}
    \item $H_{0S}$: \textit{The disagreement between raters is evenly distributed.}
    \item $H_{1S}$: \textit{The disagreement between raters is not evenly distributed.}
\end{itemize}

To test such hypotheses, we used \textbf{Bowker’s test} of internal symmetry~\citep{bowker1948test}, a non-parametric statistical test designed to evaluate whether the distribution of disagreements between two raters is symmetric across a square contingency table. In other words, it checks whether the frequency of disagreements in one direction (e.g., human says A, LLM says B) is statistically different from the reverse (LLM says A, human says B). This is particularly useful when dealing with categorical data involving more than two outcome categories, as is the case in our RM classification. Bowker’s test extends the well-known McNemar test\citep{mcnemar1947note}, which is limited to 2x2 tables, by allowing for multi-class comparisons. We selected Bowker’s test specifically due to the higher dimensionality of our data, which includes multiple motivation types, making it a more suitable and robust choice for assessing symmetry in rater disagreement.

\begin{table}
\caption{Definitions of the considered refactoring types~\cite{fowler1999refactoring} with unique ground truth motivations (\#)~\cite{silva2016we}.}
\label{tab:refactoring-motivations-related}
\footnotesize
\begin{tabular}{@{}p{0.75cm}|p{0.3cm}|p{11.6cm}@{}} 
 \toprule
 \textbf{Type}  & \textbf{\#} & \textbf{Definition} \\
 \midrule
 EM & 11 & Takes a clump of code and turns it into its method.\\  
MovC & 9 & Move the class to its new folder on the source tree. \\ 
MA & 2 & All attributes matching the selected attribute name on tags with the selected tag name may be moved inward toward a subtag of a given name. \\ 
RPack & 3 & Renames the name of the selected project package. \\ 
MM & 5 & Creates a new method with a similar body in the class that it uses the most. Either turn the old method into a simple delegation or remove it altogether. \\ 
IM & 3 & Puts the method's body into the body of its callers and removes the method. \\ 
PUM & 1 & Moves methods with identical results on subclasses to the superclass. \\ 
PUA & 1 & Moves attributes with identical results on subclasses into the superclass. \\ 
ESup & 3 & Creates a superclass and moves the common features to the superclass. \\ 
PDM & 1 & Given a field only used by some subclasses, it moves the field to those subclasses. \\
PDA & 2 & Given an attribute only used by some subclasses, it moves the attribute to those subclasses. \\  
EI & 3 & Given two classes having part of their interfaces in common, it extracts the subset into an interface. \\
 \bottomrule
\end{tabular}%

\end{table}

Finally, we were keen to investigate whether LLMs could expand the state-of-the-art by providing new motivations for previously identified ones in the reference study~\citep{silva2016we} (RQ$_2$). In this context, we prompted the LLMs with the identified RM, the ground truth RM and developer explanation, as well as the involved commit message, and tasked them to classify identified RMs as either \texttt{related} or \texttt{not related} to the ground truth motivations, and if related, if they \texttt{expanded} the motivation.

\subsubsection{Product and Process Metrics reflecting Refactoring Motivation (RQ$_3$)} 
\label{sec:RQ3-data-analysis}
This section investigates the degree to which product and process metrics can reflect developers' refactoring motivation. Consequently, we measured the feature importance and the correlation between SMs and RMs. To such end, and since RM is notorious for being a human-free text explanation of the aims and insights that lead developers to perform refactoring, we needed a categorical, encodable variable. 

Therefore, in this RQ, we (1) performed open-coding and defined refactoring motivation categories (RMCs), and, for the latter, calculated (2) feature importance and (3) correlation level for the calculated SMs.

\paragraph{\textbf{Open-Coding.}} We derived a taxonomy of RMCs from 385 identified RMs. We prompted the LLMs to perform open coding iteratively, proposing and validating categories derived from the RMs with the option to introduce new categories. Human validation was performed over a statistically significant sample (95\% confidence level, 5\% margin of error) afterwards. We provide more details on the undergone open-coding in Appendix~\ref{sec:AppendixB_3}.

\paragraph{\textbf{Feature Importance.}} 

We evaluated the influence of SMs (Table~\ref{tab:jit-metrics}) on the RMCs using Random Forest (RF)~\citep{breiman2001random} and Extreme Gradient Boosting (XGB)~\citep{Chen2016XGBoost:Systemb} machine learning models. To evaluate each SM’s influence on RMC, we ranked them by their feature importance (FI)~\citep{sammut2011encyclopedia}, a standard feature selection technique used in prior studies to measure the importance of a variable to describe the behaviour of the predicted value after fitting the model~\citep{rajbahadur2021impact, robredo2025evaluating}. We trained the models within a multi-class prediction setting to evaluate the influence of the SMs, considering the open-coded RMCs as the dependent variable.

For RF, we used two established importance measures: Mean Decrease in Accuracy (MDA) and Mean Decrease in Gini (MDG)~\citep{han2016variable, nicodemus2011stability}. MDA reflects how much the model’s accuracy drops when a specific variable is permuted—higher values suggest higher importance. MDG, on the other hand, measures how much a variable helps reduce node impurity in decision trees—again, higher scores imply greater relevance. For XGB, we relied on Information Gain (IG), which quantifies how much each variable contributes to the model relative to the total contribution of all variables~\citep{Chen2016XGBoost:Systemb}. A higher IG indicates a stronger impact.

Since feature dimensionality can influence the learning process and FI scores, we evaluated three scenarios: one using only process metrics, one using only product metrics, and one combining all SMs.

\paragraph{\textbf{Correlation Analysis.}} To investigate the statistical relationship between RMCs and SMs, we conducted a correlation analysis. As part of testing our RQ$_3$, we first assessed the distribution of the RMC data to determine the most appropriate correlation coefficient~\citep{falessi2023enhancing}.  To test the distribution data, we defined the following hypotheses:

\begin{itemize}
    \item H$_{0\mathcal{N}}$. \textit{The detected RMCs are normally distributed.}
    \item H$_{1\mathcal{N}}$. \textit{The detected RMCs are not normally distributed.}
\end{itemize}

We tested \emph{\textbf{H$_{0\mathcal{N}}$}} using Anderson-Darling (AD) test \cite{anderson1952asymptotic}. AD tests whether data points are sampled from a specific probability distribution, in this case, a normal distribution. According to Mishra et al., \cite{mishra2019descriptive}, the Shapiro-Wilk (SW) test \cite{shapiro1965analysis} would be more appropriate when using smaller datasets with fewer than 50 samples. However, our dataset is big enough to use AD, the more powerful statistical test for detecting most departures from normality \cite{stephens1974edf}.
According to the test results, we could reject the null hypothesis in all cases. Therefore, to test H$_{01,02}$, we must rely on a non-parametric test. More specificaly, we selected Spearman's $\rho$ coefficient~\citep{spearman1961proof}, which evaluates the monotonic relationship between two variables, and tests if one variable's change leads to the other variable's change, in the same direction (positive correlation) as well as in the opposite direction (negative correlation); and Kendall's $\tau$ coefficient~\citep{Wohlin2000}, which evaluates the similarity among two variables through their ordinal association. The correlation coefficient will denote a positive correlation when the observations for the pair of variables present a similar rank, and conversely, it will present a negative correlation if the orderings are dissimilar.  To interpret $\rho$ and $\tau$, we adopt Dancey and Reidy interpretation \cite{dancey2007statistics} (see Table \ref{tab:correlation-interpretation}).
 
\begin{table}[t]

\centering
\caption{Spearman’s $\rho$ and Kendall’s $\tau$ interpretation.}
\label{tab:correlation-interpretation}
\footnotesize
\begin{tabular}{@{}c|c|ccc|ccc|ccc|c@{}}
\toprule
 & \textbf{Perfect} & \multicolumn{3}{c|}{\textbf{Strong}} & \multicolumn{3}{c|}{\textbf{Moderate}} & \multicolumn{3}{c|}{\textbf{Weak}} & \textbf{Zero} \\
\midrule
\textbf{$\rho$} 
& $\pm1.0$ \hspace{-2pt}
& $\pm0.9$ & $\pm0.8$ & $\pm0.7$ \hspace{-2pt}
& $\pm0.6$ & $\pm0.5$ & $\pm0.4$ \hspace{-2pt}
& $\pm0.3$ & $\pm0.2$ & $\pm0.1$ \hspace{-2pt}
& $0$ \\
\textbf{$\tau$} 
& $\pm1.0$ \hspace{-2pt}
& $\pm0.8$ & $\pm0.7$ & $\pm0.6$ \hspace{-2pt}
& $\pm0.5$ & $\pm0.4$ & $\pm0.3$ \hspace{-2pt}
& $\pm0.2$ & $\pm0.1$ & $\pm0.05$ \hspace{-2pt}
& $0$ \\
\bottomrule
\end{tabular}%
\end{table}

Finally, since we rely on many hypothesis tests, we must mitigate the risk of Type I error or \textit{family-wise} error at the level of $\alpha$~\citep{holm1979simple}. When multiple tests are conducted simultaneously, the probability of falling into the aforementioned error increases. Therefore, we adopted the Bonferroni correction test~\citep{wright1992adjusted}, addressing the risk by adjusting globally the significance level ($\alpha'=\frac{\alpha}{m} = \frac{0.05}{574} = 0,00008711$), where $\alpha$ is the standard pre-defined significance level, i.e., usually equal to 0.05, and \textit{m} is the number of performed hypothesis operations. Nevertheless, existing literature acknowledges that Bonferroni is overly conservative when dealing with large-scale hypothesis testing, therefore we also adjust the p-value with the  Benjamini-Hochberg (BH) method~\citep{benjamini1995controlling} to control the False Discovery Rate (FDR). Unlike more conservative methods like Bonferroni correction that control the Family-Wise Error Rate (FWER), the BH method allows for more discoveries, i.e., rejections of null hypotheses, by limiting the expected proportion of false positives among those discoveries. It works by ranking p-values and comparing them to an increasing threshold based on their rank, offering a balance between discovery and error control~\citep{dalmasso2005simple}.

\section{Results}
\label{sec:Results}

This section presents the findings of our investigations and addresses the RQs:

\subsection{Refactoring Motivations: Alignment With Previous Studies (RQ$_1$)}
\label{sec:rq1-results}

To answer RQ$_1$ we analyzed a total of \textbf{758} pairs of RMs, where each pair consisted of two RMs, one from the RMs identified in our study and the other one corresponding to the RM reported by \citet{silva2016we}, both related to the same RT. Our findings focus on the alignment with the reference study~\citep{silva2016we} and the agreement with human experts, followed by an examination of the underlying RMs behind the observed disagreements. Therefore, we needed to assess the alignments based on RM pairs resulting from the same RT.

Considering the volume of retrieved RMs, applying a 95\% confidence level and a 5\% error rate, we sampled a subset of \textbf{198} pairs: \textbf{136} generated by the LRM, while \textbf{62} generated by V3 (see Section~\ref{sec:RQ1-RQ2-data-analysis}). Interestingly, V3 was found to produce more complete and reliable RMs in these fallback cases. As a result, the final composition of our \textit{corpus} of RMs was composed by \textbf{69\%} LRM-generated and \textbf{31\%} V3-generated.

To assess the degree of alignment between these LLM-generated RMs and the ones available in the reference study, we tasked the LLMs and three human experts to evaluate such alignment. In 59 cases out of 198, both the humans and the LLM judged no alignment with the reference study (``No-No'', True Negatives), while 97  out of 198 showed agreement with both judging alignment (``Yes-Yes'', True Positives). However, there were 42 cases of disagreement: in 8 instances, humans identified alignment that the LLM did not (``Yes-No'', False Positives), whereas, in 34 instances, the LLM assumed alignment where humans did not (``No-Yes'', False Positives) (see Table~\ref{tab:llmvshumancotagreement}).

The LLMs generated RMs aligned well with human judgment in most cases, showing agreement in 156 out of 198 cases ($\sim$80\%), considering the combination of the true positives and true negatives (see Table~\ref{tab:llmvshumancotagreement}). Furthermore, we measured IRA via Cohen’s kappa between LLM and humans, resulting in a moderate agreement ($\kappa=0.567$) and it was statistically significant ($p < 0.0001$) (see Table~\ref{tab:agreementHvLLM}). Moreover, according to Bowker’s symmetry test, the \textbf{disagreement was systematic} ($\chi^2 = 16.10$, $p < 0.0001$), suggesting the LLM consistently identifies different RMs concerning the human ones in terms of interpretation or overlooking. Therefore, we can affirm that there exists a statistically significant, moderate agreement between LLM and human RMs, but systematic differences in interpretation remain.

\begin{figure*}[t]
\centering
\scriptsize

\begin{minipage}[t]{0.55\textwidth}
\centering
\captionof{table}{Representativeness level for the refactoring types sampled in the manual validation for RQ$_1$.}
\label{tab:represented-ref-types}
\begin{tabular}{@{}l|rrr@{}}
\toprule
\textbf{RT} & \textbf{Representativeness} & \textbf{Agreement} & \textbf{Disagreement} \\ \midrule
EM         & 27.77\% & 30 & 25 \\
MM            & 17.17\% & 6 & 28 \\
MovC             & 13.64\% & 12 & 15 \\
MA         & 8.59\%  & 8 & 9 \\
IM          & 8.08\%  & 5 & 11 \\
PUM         & 5.05\%  & 8 & 1 \\
PDA    & 5.05\%  & 6 & 4 \\
ESup     & 3.54\%  & 4 & 2 \\
PDM       & 3.03\%  & 4 & 2 \\
RPack         & 3.03\%  & 1 & 5 \\
PUA      & 2.53\%  & 5 & 0 \\
EI      & 2.53\%  & 5 & 0 \\ \bottomrule
\end{tabular}
\end{minipage}
\hfill
\begin{minipage}[t]{0.39\textwidth}
\centering
\captionof{table}{LLM Motivation Alignment\\Agreement Contingency Table.}
\label{tab:llmvshumancotagreement}
\begin{tabular}{@{}llrr@{}}
\toprule
\textbf{LLM} & \textbf{Human} & \textbf{Frequency} & \textbf{\%} \\ \midrule
No & No   & 59 & 29.79 \\
No & Yes  & 8  & 4.04 \\
Yes & No  & 34 & 17.17 \\
Yes & Yes & 97 & 48.99 \\ \bottomrule
\end{tabular}

\vspace{0.6cm}

\captionof{table}{Human Validated LLM RM\\Alignment with reference study RMs.}
\label{tab:human-llm-alignment}
\begin{tabular}{@{}lrr@{}}
\toprule
\textbf{Classification} & \textbf{Frequency} & \textbf{\%} \\ \midrule
Extends & 44 & 22.22 \\
No      & 49 & 24.74 \\
Yes     & 105 & 53.03 \\ \bottomrule
\end{tabular}
\end{minipage}
\end{figure*}

We were also keen to investigate whether specific refactorings were more challenging to motivate and agree upon. The most frequently represented operations were EM (27.77\%), MM (17.17\%), and MC (13.64\%) (see Table~\ref{tab:represented-ref-types}). Less common types included RPack (3.03\%), PUA (2.53\%), and EI (2.53\%).

PUM showed the strongest agreement between LLM and human RMs (8 out of 9 pairs aligned), followed by PUA and EI, both with perfect agreement (5 out of 5). In contrast, MM presented the most disagreement, with 28 disagreements out of 34 comparisons. Inline Method (11 out of 16) and MC (15 out of 27) also showed significant disagreement. These results suggest that LLMs struggle more with refactorings involving method or class movement.

Based on human annotations, \textbf{97 out of 198 RMs (approximately 49\%) aligned with those reported in the reference study}.

\begin{table}
\centering
\footnotesize
\caption{Agreement Test Human Motivation Vs LLM}
\label{tab:agreementHvLLM}
\begin{tabular}{m{5.5cm}|llll}
\hline
\textbf{Agreement Test} & \textbf{Kappa} & \textbf{Std Err} & \textbf{Lower 95\%} & \textbf{Upper 95\%} \\
 & 0,567 & 0,057 & 0,455 & 0,679 \\ \hline
\textbf{Asymptotic Test} & \textbf{Prob $>$ z} & \textbf{Prob$>|Z|$} &  &  \\
 & \textit{$<,0001$} & $<,0001$ &  &  \\ \hline
\textbf{Bowker’s Test: Symmetry of Disagreement} & \textbf{ChiSquare} & \textbf{Prob$>$ChiSq} &  &  \\
 & 16,095 & $<,0001$ &  &  \\ \hline
\end{tabular}%
\end{table}

Finally, to understand the rationale between humans and the LLMs,  we manually analyzed the 42 disagreement cases, categorizing them into \textbf{five recurring disagreement categories}, each reflecting a fundamental difference in how LLMs and humans approach the interpretation of RMs (see Table~\ref{tab:disagreement-categories}). All in all, LLMs tend to reason based on localized code context, e.g., method names or syntax-level cues, likely because they lack access to full project information. In contrast, human raters apply a more holistic perspective, considering architectural structures and long-term design goals when assessing RMs.

For instance, in the Future vs Present Orientation category, humans often justify changes based on anticipated future needs, such as extensibility or scalability, whereas LLMs focus on immediate benefits, like current testability or readability. Similarly, in the Different Focus of Refactoring Granularity category, LLMs interpret changes at the method level, while humans consider higher-level structures, such as classes or packages. These patterns suggest that LLMs are proficient in identifying surface-level improvements, but they frequently miss deeper, strategic rationales that require an understanding of broader software design intent.

\begin{table}
\centering
\caption{Summary of manually detected disagreement categories between LLM extracted motivations and those from the reference study.}
\label{tab:disagreement-categories}
\footnotesize
\begin{tabular}{@{}p{0.35\linewidth}|p{0.55\linewidth}@{}}
\hline
\textbf{Disagreement category}                         & \textbf{Description}                                                                                                                                                               \\ \hline
\textbf{Different Focus of Refactoring Granularity}    & Human focuses on attributes, classes, or packages while LLM focuses more localized structures such as methods.                                                                     \\
\textbf{Intent Misalignment: Structural vs Functional} & Human focuses on structural aspects (e.g. clarity, visibility...) while LLM emphasizes on a more functional perspective (e.g. testability...)                                      \\
\textbf{Future vs Present Orientation}                 & Humans often refer to future needs in their motivations (e.g. extension, scalability), while LLM focuses on immediate operations such as current testing needs.                    \\
\textbf{Interpretation of Refactoring Scope}           & Human views refactoring as modular reorganization while LLM sees the operation as an isolated change.                                                                              \\
\textbf{Semantic vs Syntactic Understanding}           & Human centres on semantic changes to clarify intentional code statements (e.g. ownership, package naming...) while LLM refers to class organisation or simple syntactical clarity. \\ \hline
\end{tabular}%
\end{table}

\begin{keyTakeAways}[Aligned or not Aligned?]
Only \textbf{47\% of the LLM-generated RMs matched prior reference study} according to human evaluation. While there was an \textbf{80\% agreement between LLMs and humans overall}, systematic differences emerged; LLMs rely heavily on \textbf{local code cues}, while humans factor in \textbf{architectural and strategic goals}. Certain refactorings like \textbf{Pull Up Method} saw strong alignment, but \textbf{Move Method} and \textbf{Move Class} revealed persistent LLM struggles, emphasizing the need for deeper context understanding to align fully with past studies.

\end{keyTakeAways}

\subsection{Refactoring Motivations: Extensions of Previous Studies (RQ$_2$)}

To investigate how LLM-generated motivation content compares with prior studies, we examined the degree to which they extend the state of the art. 
In \textbf{105 out of 198 analyzed cases (53\% )} showed direct matches with RMs previously reported in the reference study (see Table~\ref{tab:human-llm-alignment}).  Moreover, in \textbf{44 cases out of 198 (22\%)}, the LLMs also \textbf{agreed} on the motivation reported in the reference study, but they also inferred \textbf{meaningful extensions or refinements} with information such as context-specific or fine-grained justifications.

Conversely, \textbf{49 cases (25\%)} did \textbf{not align with any existing motivation}, hinting at \textbf{potentially novel rationales}.

More specifically, we identified seven distinct rationales on how LLM-generated RMs expanded human-reported ones (see Table~\ref{tab:LLM_extensions_summary}). These include adding missing technical details,e.g., clarifying the benefits of a specific refactoring operation that human RMs overlooked, expanding the scope of changes, e.g., covering both method and attribute moves, and emphasizing aspects such as readability, structural clarity, and testability.

\begin{table}
\centering
\caption{Summary of LLM Motivations that Extend Human-Reported Motivations.}
\label{tab:LLM_extensions_summary}
\resizebox{0.95\linewidth}{!}{%
\footnotesize
\begin{tabular}{p{5.3cm}|p{8cm}}
\hline
\textbf{Characteristics of LLM Extensions} & \textbf{Example Description} \\ 
\hline
\textbf{Enhanced Detail and Precision} & The LLM motivation clarifies human-provided motivations with more precise and detailed reasoning. \\[4pt]

\textbf{Broader Scope} & LLM identifies additional aspects of refactoring (e.g., moving attributes in addition to methods). \\[4pt]

\textbf{Maintainability and Readability Emphasis} & LLM highlights code maintainability and readability, extending beyond human focus on code functionality. \\[4pt]

\textbf{Structural Clarity} & LLM explicitly addresses improvements to the structural clarity and dependency injection of the code. \\[4pt]

\textbf{Comprehensive Renaming Context} & Human motivation focuses narrowly on renaming; LLM elaborates underlying reasons for renaming. \\[4pt]

\textbf{Improved Explanation} & The LLM provides clearer and more explicit explanations of underlying developer motivations. \\[4pt]

\textbf{Explicit Testing and Flexibility Context} & LLM explicitly includes motivations related to improving testing and code flexibility. \\
\hline
\end{tabular}%
}
\end{table}

\begin{keyTakeAways}[LLMs extend the state of the art]
While \textbf{53\%} of the RMs matched known reasons from past studies, the remaining \textbf{47\%} revealed either \textbf{extended} (22\%) or \textbf{entirely new} (25\%) RMs. These included overlooked technical details, broader refactoring scopes, and added focus on \textbf{readability, structure, and testability}, highlighting that developers may be driven by \textbf{richer and more diverse RMs} than previously collected.

\end{keyTakeAways}

\subsection{Metrics and Refactoring Motivations (RQ$_3$)}

This section investigates the relationship between developer RMs for refactoring and SMs. Our aim is threefold: (1) to categorize the RMs extracted from LLMs via open coding, (2) to rank the SMs based on their ability to discriminate among RMCs, and (3) to analyze the statistical correlations between the identified RMCs and the computed metrics.

\subsubsection{Open-Coding Refactoring Motivations into Categories}

We first conducted open coding on 385 sampled refactoring observations to identify recurring RMs. This process yielded \textbf{167 distinct RMs}, which we grouped into \textbf{14 RMCs} (see Table~\ref{tab:motivation-categories-families}). These categories encompass common refactoring rationales such as \textbf{Code Clarity and Readability (CCR)}, \textbf{Code Simplification and Redundancy Reduction (CSRR)}, and more specialized goals like \textbf{Encapsulation and Abstraction}.

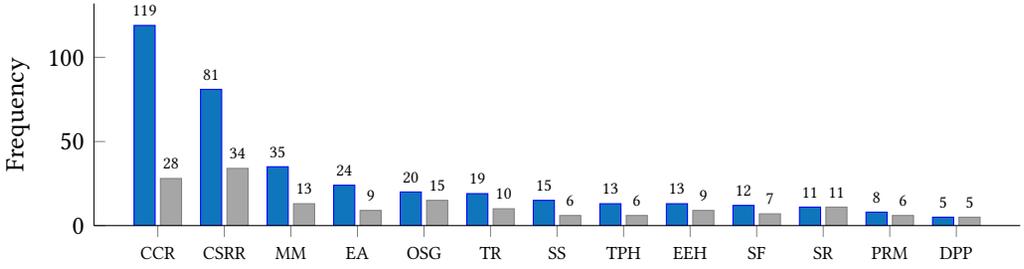
\begin{figure}[t]
\centering
\begin{tikzpicture}
\begin{axis}[
    ybar,
    bar width=8pt,
    width=\textwidth,
    height=4.5cm,
    enlarge x limits=0.08,
    ylabel={Frequency},
    symbolic x coords={CCR,CSRR,MM,EA,OSG,TR,SS,TPH,EEH,SF,SR,PRM,DPP},
    xtick=data,
    x tick label style={font=\scriptsize},
    nodes near coords,
    nodes near coords align={vertical},
    ymin=0,
    ymax=120,
    enlarge y limits={upper,value=0.1},
    axis lines*=left,
    axis on top,
    every node near coord/.append style={font=\tiny, color=black}
]
\addplot+[fill=acm-blue, draw=blue] coordinates {
    (CCR,119) (CSRR,81) (MM,35) (EA,24) (OSG,20)
    (TR,19) (SS,15) (TPH,13) (EEH,13) (SF,12)
    (SR,11) (PRM,8) (DPP,5)
};

\addplot+[fill=gray!70, draw=gray] coordinates {
    (CCR,28) (CSRR,34) (MM,13) (EA,9) (OSG,15)
    (TR,10) (SS,6) (TPH,6) (EEH,9) (SF,7)
    (SR,11) (PRM,6) (DPP,5)
};
\end{axis}
\end{tikzpicture}
\caption{Frequencies for extracted total motivations (\textbf{blue}) and unique motivation instances (\textbf{grey}).}
\label{fig:rmc-frequencies}
\end{figure}

Figure~\ref{fig:rmc-frequencies} summarizes the frequency distribution of the extracted RMs. Among all, \textbf{CCR} (30.91\%) and \textbf{CSRR} (24.04\%) were the most frequently occurring RMCs. However, when considering only the unique instances, CSRR (20.36\%) slightly surpassed CCR (17.36\%), indicating a richer diversity of simplification-related RMs.

Among the most dominant RMCs, \textbf{Code Clarity and Readability (CCR)} emerged as the most frequent, with 119 occurrences. This category encapsulates RMs centred on enhancing the comprehensibility, abstraction, and ease of reading code. It includes cases where developers aim to refactor by renaming variables for clarity, extracting methods to reduce cognitive load, or organizing code to align with human-readable logic.

Following closely, \textbf{Code Simplification and Redundancy Reduction (CSRR)} appeared 81 times and reflects RMs that emphasize minimizing unnecessary complexity. This includes reducing duplicated logic, collapsing verbose constructs, or eliminating parameters and variables that no longer serve a purpose. Together, CCR and CSRR account for over half of the observed RMs, reinforcing that LLM-generated RMs often prioritize the legibility and minimalism of code.

The third most common category, \textbf{Maintainability and Modularity (MM)} (35 occurrences), reflects efforts to improve the long-term evolvability and structural organization of the codebase. RMs in this category often involve modularizing components, encapsulating change-prone logic, and enhancing the separation of concerns to support sustainable maintenance.

Beyond these primary themes, several RMCs address more specific technical aspects. For example, \textbf{Encapsulation and Abstraction (EA)} (24) focuses on isolating responsibilities and reducing the surface of external interactions, which includes practices like hiding implementation details or reducing class coupling. Similarly, \textbf{Testing and Reliability (TR)} (19) relates to improving the testability of code, often through simplifying logic or making control flows more deterministic.

A notable portion of RMs also fell into \textbf{Other Specialized Goals (OSG)} (20), which act as a catch-all for domain-specific or niche objectives that do not neatly fit into other categories. This includes specialized algorithmic refactoring, compliance with domain-specific constraints, or integration-related concerns. Less common but still essential RMCs included:

\begin{itemize}
    \item \textbf{Security and Safety (SS)} (15), addressing concerns like thread safety, null safety, or eliminating dangerous constructs.
    \item \textbf{Exception and Error Handling (EEH)} (13), which involves improving how code deals with failures or unexpected states.
    \item \textbf{Type and Parameter Handling (TPH)} (13), focusing on method signatures, type safety, and semantic correctness.
    \item \textbf{Support for (New) Functionalities (SF)} (12), where the motivation is driven by expanding or improving system capabilities.
    \item \textbf{Structural Reorganization (SR)} (11), representing architectural restructuring actions like moving classes or extracting responsibilities.
    \item \textbf{Consistency and Standardization (CS)} (10), which targets uniformity in naming, formatting, or style to align with project standards.
    \item \textbf{Performance and Resource Management (PRM)} (8), dealing with optimizations such as memory usage, threading, or execution time.
    \item \textbf{Design Principles and Patterns (DPP)} (5), the least represented, where the motivation stems from aligning code with well-known design patterns or architectural principles.
\end{itemize}

Such a taxonomy of RMs lays the foundation for subsequent analysis, allowing us to study how measurable SMs correlate with or indicate the presence of specific RMCs.

\begin{table}[t]
\tiny
\centering
\caption{Description of the open-coded refactoring motivation categories (RMC).}
\label{tab:motivation-categories-families}
\resizebox{\linewidth}{!}{%
\begin{tabular}{m{0.15\linewidth}m{0.35\linewidth}m{0.01\linewidth}m{0.2\linewidth}}
\hline
\textbf{Motivation Category} & \textbf{Description} & \textbf{Occurrences}  & \textbf{} \\ \hline 
\raggedright Code Clarity and Readability (CCR) & Motivations aiming to improve the readability, abstraction, and understandability of the code. & 119 & \sbar{119}{119} \\
\raggedright Code Simplification and Redundancy Reduction (CSRR) & Focus on reducing complexity, eliminating duplication, and streamlining code. & 81 & \sbar{81}{119} \\
\raggedright Maintainability and Modularity (MM) & Focuses on long-term maintainability and modular decomposition of software components. & 35 & \sbar{35}{119} \\
\raggedright Encapsulation and Abstraction (EA) & Deals with isolating responsibilities and minimizing external dependencies or access. & 24 & \sbar{24}{119} \\
\raggedright Other Specialized Goals (OSG) & All other motivations that serve niche, technical, or domain-specific purposes. & 20 & \sbar{20}{119} \\
\raggedright Testing and Reliability (TR) & Refactorings aimed at improving code testability and reliability. & 19 & \sbar{19}{119} \\
\raggedright Security and Safety (SS) & Motivations ensuring safer, more secure code, such as null safety or thread safety. & 15 & \sbar{15}{119} \\
\raggedright Exception and Error Handling (EEH) & Related to improving how exceptions and errors are managed in the codebase. & 13 & \sbar{13}{119} \\
\raggedright Type and Parameter Handling (TPH) & Deals with type safety, parameter handling, and semantic correctness of method inputs. & 13 & \sbar{13}{119} \\
\raggedright Support (New) Functionalities (SF) & Motivations enhancing current code functionality and introduction of new functionalities. & 12 & \sbar{12}{119} \\
\raggedright Structural Reorganization (SR) & Code transformations involving movement or reclassification of structural elements. & 11 & \sbar{11}{119} \\
\raggedright Consistency and Standardization (CS) & Focus on aligning code with standards or maintaining consistent patterns. & 10 & \sbar{10}{119} \\
\raggedright Performance and Resource Management (PRM) & Improvements targeting efficiency, memory, or threading concerns. & 8 & \sbar{8}{119} \\
\raggedright Design Principles and Patterns (DPP) & Encourages use of standard design patterns and separation of concerns. & 5 & \sbar{5}{119} \\
\hline 
\end{tabular}%
}
\end{table}

\begin{figure*}[ht]
\centering
\scriptsize
\begin{minipage}[t]{0.48\textwidth}
\centering
\captionof{table}{Refactoring Types Frequencies (\#) for CCR RMC}
\label{tab:ccr-refactoring-sorted}
\begin{tabular}{lr|lr|lr|lr}
\hline
\textbf{RT} & \textbf{\#} & \textbf{RT} & \textbf{\#} & \textbf{RT} & \textbf{\#} & \textbf{RT} & \textbf{\#} \\ \hline
EV   & 8 & EAMM & 6 & RM   & 5 & EM   & 4 \\
RC   & 4 & RV   & 4 & CRT  & 3 & RParam & 3 \\
MovC & 3 & ACM  & 3 & RAWL & 3 & AMA  & 3 \\
RLWP & 3 & IM   & 2 & MA   & 2 & CPT  & 2 \\
RCA  & 2 & MAIM & 2 & RPack & 2 & RPWL & 2 \\
RemP & 2 & MARM & 2 & MARC & 2 & SM   & 2 \\
IV   & 2 & CVT  & 2 & RVM  & 2 & MPack & 2 \\
RMA  & 2 & MAA  & 2 & RA   & 2 & AVM  & 2 \\
RPM  & 2 & ACA  & 1 & RenP & 1 & MARA & 1 \\
RCWT & 1 & SA   & 1 & TWR  & 1 & MovM & 1 \\
RA   & 1 & MerA & 1 & ESub & 1 & AAM  & 1 \\
MCon & 1 & RGWD & 1 & RAA  & 1 & RVA  & 1 \\
RPA  & 1 & MCode & 1 & SClass & 1 & CAT & 1 \\
PDA  & 1 & LP   & 1 & RCM  & 1 & RAM  & 1 \\
\hline
\multicolumn{7}{l}{(\textbf{RT}: Refactoring Type)} &
\end{tabular}
\end{minipage}
\hfill
\begin{minipage}[t]{0.48\textwidth}
\centering
\captionof{table}{Refactoring Types Frequencies (\#) for CSRR RMC}
\label{tab:csrr-refactoring-freq-sorted}
\begin{tabular}{lr|lr|lr|lr}
\hline
\textbf{RT} & \textbf{\#} & \textbf{RT} & \textbf{\#} & \textbf{RT} & \textbf{\#} & \textbf{RT} & \textbf{\#} \\ \hline
EV    & 7 & RemP  & 7 & IV    & 5 & MParam & 3 \\
MV    & 3 & PUM   & 3 & RPA   & 3 & RVM    & 3 \\
MAIM  & 2 & MerC  & 2 & MerM  & 2 & RCA    & 2 \\
RMA   & 2 & RTET  & 2 & RVA   & 2 & RLWP   & 2 \\
SC    & 2 & CAT   & 1 & CCAM  & 1 & CPT    & 1 \\
CRT   & 1 & CTDK  & 1 & ExA   & 1 & EM     & 1 \\
ESup  & 1 & IA    & 1 & IM    & 1 & IC     & 1 \\
LP    & 1 & Mcat  & 1 & MCon  & 1 & MCA    & 1 \\
MVA   & 1 & PV    & 1 & PUA   & 1 & RAA    & 1 \\
RCM   & 1 & RPM   & 1 & RAWV  & 1 & RCWT   & 1 \\
RGWD  & 1 & RVWA  & 1 & SP    & 1 &        &   \\
\hline
\multicolumn{7}{l}{(\textbf{RT}: Refactoring Type)} &
\end{tabular}
\end{minipage}
\end{figure*}

\subsubsection{Investigating Metrics Importance for Refactoring Motivations Categories}
Understanding what drives developers to initiate refactoring activities is crucial for building intelligent, context-aware recommendation systems, for instance. In this final RQ, we investigate which SMs, both process and product metrics, are most indicative of specific RMCs. We employ two state-of-the-art machine learning models, Random Forest (RF) and Extreme Gradient Boosting (XGB), to compute feature importance scores and isolate the metrics that most strongly influence model predictions, and two statistical tests, Spearman and Kendall, to measure the correlation of such metrics with the RMCs.

\paragraph{RF Results.} 
We analyzed \textbf{process metrics} importance scores according to RF in terms of Mean Decrease in Accuracy (MDA) and Mean Decrease in Gini (MDG). MDA and MDG measure how much each metric contributes to improving the model’s predictive performance (see Section~\ref{sec:RQ3-data-analysis}). 

\begin{figure}[t]
    \centering
    \begin{adjustbox}{width=1\textwidth}
\begin{tikzpicture}

\begin{axis}[
    ybar,
    bar width=10pt,
    width=15cm,
    height=4.5cm,
    ymin=0,
    ymax=8,
    ylabel={Mean decrease accuracy},
    symbolic x coords={
        ADD,ADEV,COMM,DDEV,DELE,EXP,MINOR,NADEV,NCOMM,NDDEV,NSCTR,OEXP,OWN
    },
    xtick=data,
    xticklabel style={font=\scriptsize, rotate=45, anchor=east},
    yticklabel style={font=\scriptsize},
    enlarge x limits=0.05,
    axis x line*=bottom,
    axis y line*=left,
]

\addplot+[
    ybar,
    fill=acm-blue,
] coordinates {
    (ADD,0.68980145)
    (ADEV,7.37629894)
    (COMM,3.82622743)
    (DDEV,5.30040880)
    (DELE,6.30487670)
    (EXP,3.04075209)
    (MINOR,6.70717319)
    (NADEV,1.70311761)
    (NCOMM,0)  %
    (NDDEV,0)  %
    (NSCTR,3.91497810)
    (OEXP,5.48656542)
    (OWN,2.13123542)
};

\end{axis}

\begin{axis}[
    ybar,
    ymin=0,
    bar width=10pt,
    at={(0,-1cm)},
    anchor=north west,
    width=15cm,
    height=4.5cm,
    ylabel={Mean decrease Gini},
    xlabel={},
    symbolic x coords={
        ADD,ADEV,COMM,DDEV,DELE,EXP,MINOR,NADEV,NCOMM,NDDEV,NSCTR,OEXP,OWN
    },
    xtick=data,
    xticklabel style={font=\scriptsize, rotate=45, anchor=east},
    yticklabel style={font=\scriptsize},
    enlarge x limits=0.05,
    axis x line*=bottom,
    axis y line*=left,
]

\addplot+[
    ybar,
    fill=gray!70,
    draw=gray,
] coordinates {
    (ADD,35.080336)
    (ADEV,22.466547)
    (COMM,36.355737)
    (DDEV,22.557860)
    (DELE,34.627562)
    (EXP,38.369866)
    (MINOR,18.013449)
    (NADEV,8.705541)
    (NCOMM,8.755858)
    (NDDEV,6.300576)
    (NSCTR,35.946236)
    (OEXP,38.281008)
    (OWN,5.520730)
};

\end{axis}
\end{tikzpicture}
\end{adjustbox}
    \vspace{-2.2em}
    \caption{MDA and MDG importance results for Random Forest model trained with process metrics.}
    \label{fig:randomForestImportance-process}
\end{figure}
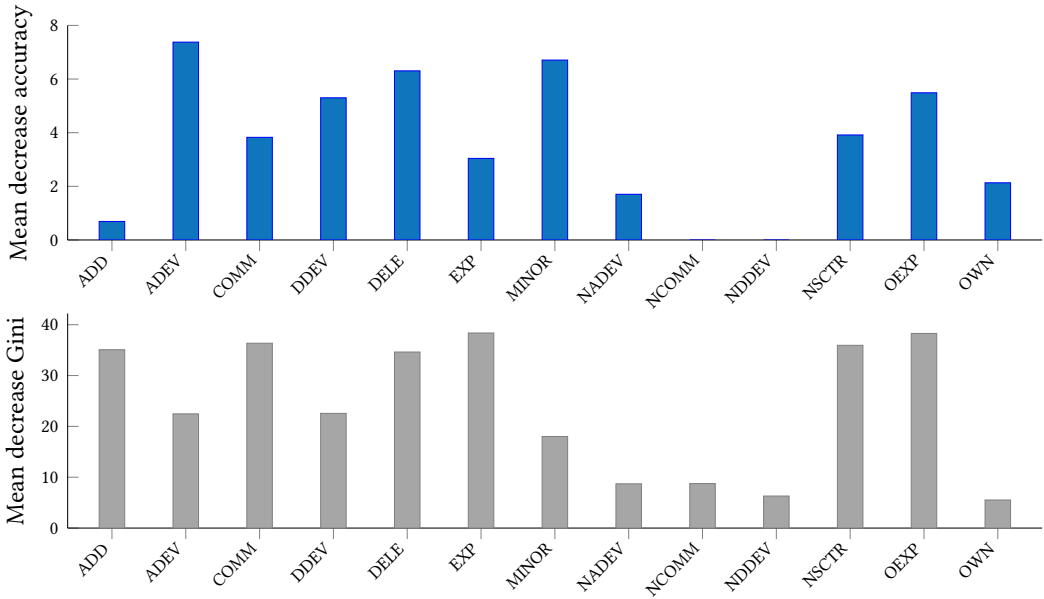

From the \textbf{MDA perspective}, which reflects how much the model’s accuracy decreases when a feature is randomly permuted, the most informative process metrics were \textbf{ADEV} (the cumulative number of active developers per file) with a score of 7.37, and \textbf{MINOR} (number of minor contributors to a file) with 6.70, hinting that the diversity and activity of contributors are particularly relevant signals when trying to predict the target outcome in the model (see Figure~\ref{fig:randomForestImportance-process}).

Conversely, \textbf{MDG}, which reflects how each variable contributes to the purity of the decision tree splits (i.e., how well it separates the data), top-ranked  \textbf{EXP} (average developer experience in the project, 38.37), \textbf{OEXP} (ownership experience on a file, 38.28), \textbf{COMM} (number of commits per file, 36.35), and \textbf{NSCTR} (developer spread across packages, 35.94). Hence, we note that long-term developer familiarity with the project and the intensity of change activity in files and packages are key indicators used by the model to make predictions.

All in all, these findings reveal that both \textbf{developer-related characteristics} (e.g., experience and ownership) and \textbf{collaboration dynamics} (e.g., number and type of contributors) play a central role in shaping the model’s understanding of the data.

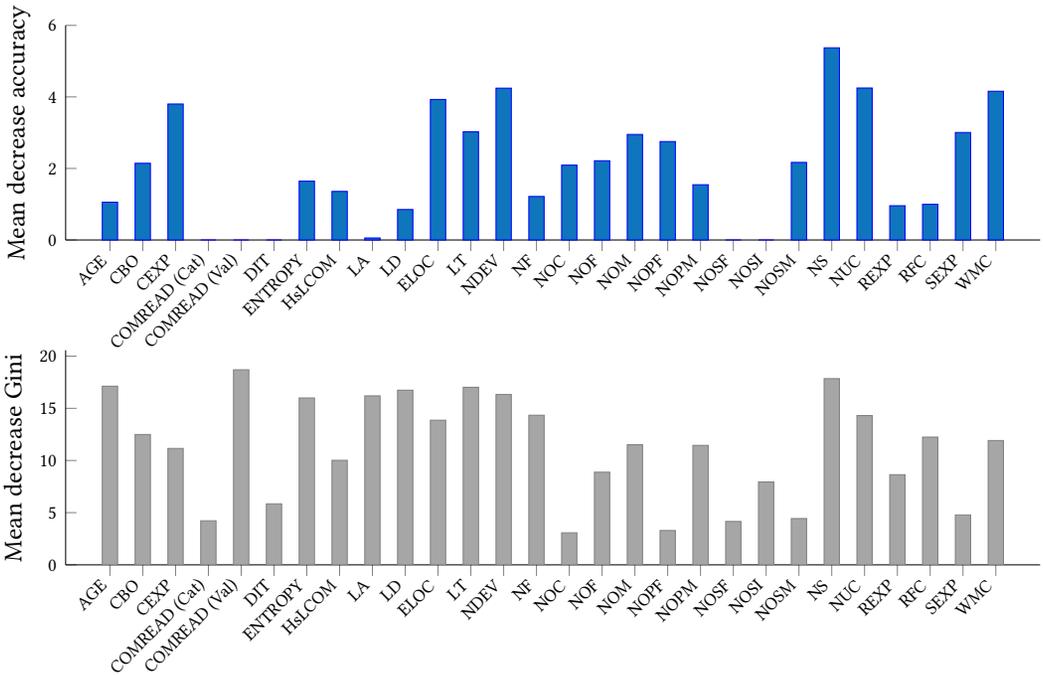
\begin{figure}[t]
    \centering
    \begin{adjustbox}{width=1\textwidth}
\begin{tikzpicture}

\begin{axis}[
    ybar,
    bar width=6pt,
    width=15cm,
    height=4.5cm,
    ymin=0,
    ymax=6,
    ylabel={Mean decrease accuracy},
    symbolic x coords={
        AGE,CBO,CEXP,COMREAD (Cat),COMREAD (Val),DIT,ENTROPY,HsLCOM,LA,LD,ELOC,
        LT,NDEV,NF,NOC,NOF,NOM,NOPF,NOPM,NOSF,NOSI,NOSM,NS,NUC,REXP,RFC,SEXP,WMC
    },
    xtick=data,
    xticklabel style={font=\scriptsize, rotate=45, anchor=east},
    yticklabel style={font=\scriptsize},
    enlarge x limits=0.05,
    axis x line*=bottom,
    axis y line*=left,
]

\addplot+[
    ybar,
    fill=acm-blue,
] coordinates {
    (AGE,1.05638880)
    (CBO,2.14434399)
    (CEXP,3.80090970)
    (COMREAD (Cat),0)  %
    (COMREAD (Val),0)  %
    (DIT,0)            %
    (ENTROPY,1.64446827)
    (HsLCOM,1.35774727)
    (LA,0.05623184)
    (LD,0.85086458)
    (ELOC,3.92867933)
    (LT,3.02444168)
    (NDEV,4.24463091)
    (NF,1.21544901)
    (NOC,2.09210148)
    (NOF,2.21507301)
    (NOM,2.95087753)
    (NOPF,2.75195779)
    (NOPM,1.54451979)
    (NOSF,0)           %
    (NOSI,0)           %
    (NOSM,2.16905084)
    (NS,5.37237919)
    (NUC,4.24930744)
    (REXP,0.95718789)
    (RFC,0.99717828)
    (SEXP,3.00495890)
    (WMC,4.15790064)
};

\end{axis}

\begin{axis}[
    ybar,
    ymin=0,
    bar width=6pt,
    at={(0,-1.5cm)},
    anchor=north west,
    width=15cm,
    height=4.5cm,
    ylabel={Mean decrease Gini},
    symbolic x coords={
        AGE,CBO,CEXP,COMREAD (Cat),COMREAD (Val),DIT,ENTROPY,HsLCOM,LA,LD,ELOC,
        LT,NDEV,NF,NOC,NOF,NOM,NOPF,NOPM,NOSF,NOSI,NOSM,NS,NUC,REXP,RFC,SEXP,WMC
    },
    xtick=data,
    xticklabel style={font=\scriptsize, rotate=45, anchor=east},
    yticklabel style={font=\scriptsize},
    enlarge x limits=0.05,
    axis x line*=bottom,
    axis y line*=left,
]

\addplot+[
    ybar,
    fill=gray!70, 
    draw=gray,
] coordinates {
    (AGE,17.117806)
    (CBO,12.488329)
    (CEXP,11.153319)
    (COMREAD (Cat),4.220924)
    (COMREAD (Val),18.691372)
    (DIT,5.845893)
    (ENTROPY,15.990738)
    (HsLCOM,10.015390)
    (LA,16.192397)
    (LD,16.735048)
    (ELOC,13.858839)
    (LT,17.013890)
    (NDEV,16.322018)
    (NF,14.335426)
    (NOC,3.068147)
    (NOF,8.876766)
    (NOM,11.510846)
    (NOPF,3.288878)
    (NOPM,11.445427)
    (NOSF,4.156310)
    (NOSI,7.940645)
    (NOSM,4.434768)
    (NS,17.839624)
    (NUC,14.303096)
    (REXP,8.637583)
    (RFC,12.238719)
    (SEXP,4.783790)
    (WMC,11.910337)
};

\end{axis}
\end{tikzpicture}
\end{adjustbox}
    \vspace{-2.2em}
    \caption{MDA and MDG importance results for Random Forest model trained with product metrics.}
    \label{fig:randomForestImportance-product}
\end{figure}

Regarding the importance of \textbf{product metrics}, from the \textbf{MDA} perspective, the top-ranking product metrics were \textbf{NS} (number of modified subsystems, 5.37) and \textbf{NUC} (number of file changes up to the current commit, 4.25). Such metrics reflect structural and evolutionary dimensions of the system, i.e., \textbf{NS} reflects the scope of change across architectural units (i.e., subsystems), while \textbf{NUC} emphasizes how frequently a file is subject to modification (see Figure~\ref{fig:randomForestImportance-product}). Their high importance suggests that the \textbf{extent of system modularity} and \textbf{change-proneness of code artifacts} are critical indicators for identifying relevant behavioral patterns such as RMC (Refactoring- or Maintenance-related Changes).

On the other hand, the \textbf{MDG} scores point to a different set of influential factors. The most prominent metric was \textbf{COMREAD (Val)} (18.69), which assesses the \textbf{comprehensive readability} of a class by incorporating textual, structural, and visual cues. This was followed by \textbf{NS} (17.84), \textbf{AGE} (average time since the last change, 17.12), \textbf{LT} (lines of code before the change, 17.01), and \textbf{LD} (lines deleted, 16.73). These findings underline the role of \textbf{code readability, code age, and size-related factors} in shaping the model’s predictions. More specifically, the high relevance of \textbf{readability} suggests that the ease of understanding code, possibly influencing developer decisions, could be a key factor in detecting RMCs.
All in all, the differences between MDA and MDG emphasize that while \textbf{structural evolution and change intensity} dominate in MDA-based importance, \textbf{code quality and maintainability dimensions} such as readability and code churn gain prominence when viewed through the lens of MDG.

\begin{figure}[t]
    \centering
    \begin{adjustbox}{width=1\textwidth}
\begin{tikzpicture}

\begin{axis}[
    ybar,
    bar width=6pt,
    width=18cm,
    height=4.5cm,
    ymin=0,
    ymax=6,
    ylabel={Mean decrease accuracy},
    symbolic x coords={
        ADD,ADEV,AGE,CBO,CEXP,COMM,COMREAD (Cat),COMREAD (Val),DDEV,DELE,DIT,
        ENTROPY,EXP,FIX,HsLCOM,LA,LD,ELOC,LT,MINOR,NADEV,NCOMM,NDDEV,NDEV,NF,
        NOC,NOF,NOM,NOPF,NOPM,NOSF,NOSI,NOSM,NS,NSCTR,NUC,OEXP,OWN,REXP,RFC,
        SEXP,WMC
    },
    xtick=data,
    xticklabel style={font=\scriptsize, rotate=45, anchor=east},
    yticklabel style={font=\scriptsize},
    enlarge x limits=0.02,
    axis x line*=bottom,
    axis y line*=left,
]

\addplot+[
    ybar,
    fill=acm-blue,
] coordinates {
    (ADD,2.39) (ADEV,4.13) (AGE,0.68) (CBO,0.49) (CEXP,1.64) (COMM,1.67)
    (COMREAD (Cat),2.09) (COMREAD (Val),0.56) (DDEV,2.71) (DELE,2.98) (DIT,0.00)
    (ENTROPY,0.00) (EXP,0.32) (FIX,0.00) (HsLCOM,0.00) (LA,0.00) (LD,1.26)
    (ELOC,3.05) (LT,2.50) (MINOR,4.64) (NADEV,0.32) (NCOMM,0.00)
    (NDDEV,0.51) (NDEV,4.34) (NF,2.07) (NOC,1.02) (NOF,2.11) (NOM,2.75)
    (NOPF,0.00) (NOPM,0.00) (NOSF,0.04) (NOSI,0.00) (NOSM,0.07)
    (NS,5.47) (NSCTR,3.75) (NUC,2.34) (OEXP,3.44) (OWN,2.60) (REXP,1.60)
    (RFC,2.05) (SEXP,5.65) (WMC,3.05)
};

\end{axis}

\begin{axis}[
    ybar,
    ymin=0,
    bar width=6pt,
    at={(0,-1.5cm)},
    anchor=north west,
    width=18cm,
    height=4.5cm,
    ylabel={Mean decrease Gini},
    xlabel={},
    symbolic x coords={
        ADD,ADEV,AGE,CBO,CEXP,COMM,COMREAD (Cat),COMREAD (Val),DDEV,DELE,DIT,
        ENTROPY,EXP,FIX,HsLCOM,LA,LD,ELOC,LT,MINOR,NADEV,NCOMM,NDDEV,NDEV,NF,
        NOC,NOF,NOM,NOPF,NOPM,NOSF,NOSI,NOSM,NS,NSCTR,NUC,OEXP,OWN,REXP,RFC,
        SEXP,WMC
    },
    xtick=data,
    xticklabel style={font=\scriptsize, rotate=45, anchor=east},
    yticklabel style={font=\scriptsize},
    enlarge x limits=0.02,
    axis x line*=bottom,
    axis y line*=left,
]

\addplot+[
    ybar,
    fill=gray!70, 
    draw=gray,
] coordinates {
    (ADD,11.10) (ADEV,7.35) (AGE,11.28) (CBO,8.56) (CEXP,7.56) (COMM,11.74)
    (COMREAD (Cat),2.93) (COMREAD (Val),12.49) (DDEV,7.38) (DELE,10.63)
    (DIT,3.90) (ENTROPY,10.81) (EXP,13.53) (FIX,0.00) (HsLCOM,6.83)
    (LA,10.73) (LD,10.85) (ELOC,8.76) (LT,11.13) (MINOR,5.83)
    (NADEV,3.34) (NCOMM,3.31) (NDDEV,2.79) (NDEV,10.65) (NF,9.63)
    (NOC,2.00) (NOF,6.16) (NOM,7.44) (NOPF,2.46) (NOPM,8.00) (NOSF,3.01)
    (NOSI,5.61) (NOSM,2.96) (NS,11.97) (NSCTR,11.97) (NUC,8.87)
    (OEXP,13.84) (OWN,1.78) (REXP,5.52) (RFC,7.96) (SEXP,3.59) (WMC,8.18)
};

\end{axis}
\end{tikzpicture}
\end{adjustbox}
    \vspace{-2.2em}
    \caption{MDA and MDG importance results for Random Forest model trained with process and product metrics.}
    \label{fig:randomForestImportance}
\end{figure}
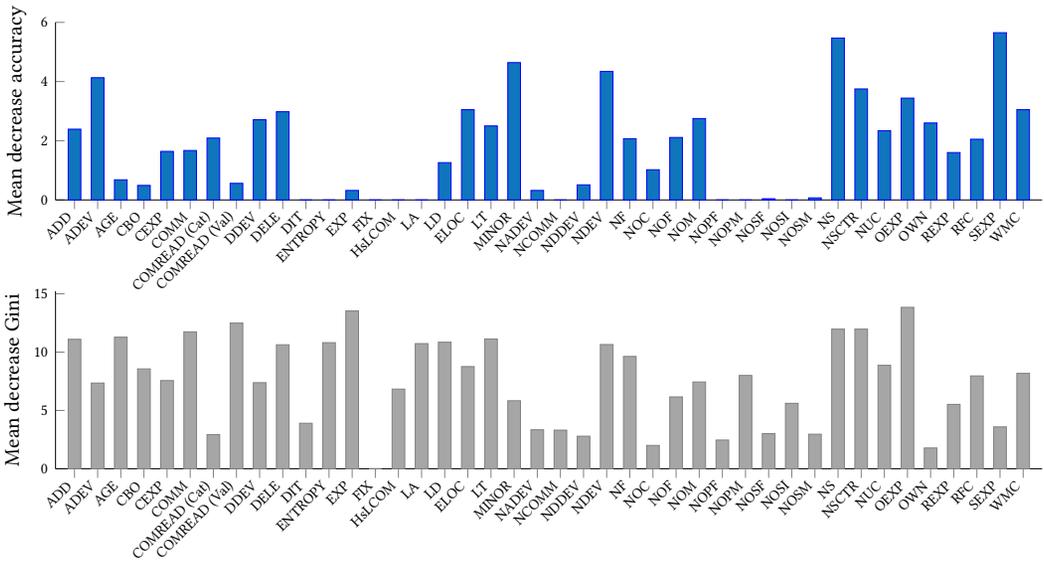

Finally, regarding the \textbf{combined importance of both product and process metrics}, \textbf{MDA results} emphasize the influence of developer-related characteristics, with \textbf{SEXP} and \textbf{NS} emerging as the most influential metrics (scores of 5.65 and 5.47, respectively). These are followed closely by \textbf{MINOR}, \textbf{NDEV}, and \textbf{ADEV}, further underscoring the significance of both developers' \textbf{experience} and \textbf{collaboration dynamics} in predicting refactoring, or main developers-related changes (RMCs) (see Figure~\ref{fig:randomForestImportance}). This blend of human and structural factors suggests that both the \textit{quantity} and \textit{diversity} of developer contributions shape the likelihood of code evolution events.
On the other hand, the \textbf{MDG scores} highlight another important layer of insight. The top metrics, \textbf{OEXP} (13.84), \textbf{EXP} (13.53), and \textbf{COMREAD} (12.49), collectively represent \textbf{developer ownership}, \textbf{project experience}, and \textbf{code comprehensibility}. These were followed closely by \textbf{NS} and \textbf{NSCTR} (both 11.97), showing that the structural breadth of change across subsystems and packages also plays a key role. Altogether, the MDG-based ranking complements the MDA findings, reinforcing the conclusion that RMCs are influenced by a \textbf{combination of code quality, developer engagement, and structural complexity}.

\begin{figure}[t]
    \centering
    \begin{adjustbox}{width=1\textwidth}
\begin{tikzpicture}
\begin{axis}[
    ybar,
    bar width=10pt,
    height=4.5cm,
    width=14cm,
    ymin=0,
    ymax=0.18,
    ylabel={Importance (Gain)},
    xtick=data,
    xticklabel style={
        rotate=45,
        anchor=east,
        font=\scriptsize
    },
    yticklabel style={
        font=\scriptsize
    },
    symbolic x coords={
        NSCTR, EXP, COMM, OEXP, ADD,
        DELE, ADEV, DDEV, MINOR, NADEV,
        OWN (False), NDDEV
    },
    axis lines*=left,
    enlarge x limits=0.05,
    point meta=y,
    title={},
]

\addplot+[ybar, fill=acm-blue, draw=blue] coordinates {
    (NSCTR,     0.160555464)
    (EXP,       0.153807161)
    (COMM,      0.137980212)
    (OEXP,      0.137237657)
    (ADD,       0.131833603)
    (DELE,      0.101948085)
    (ADEV,      0.071341071)
    (DDEV,      0.045222895)
    (MINOR,     0.025931538)
    (NADEV,     0.021579195)
    (OWN (False),  0.008065597)
    (NDDEV,     0.004497523)
};

\end{axis}
\end{tikzpicture}
\end{adjustbox}
    \vspace{-2.2em}
    \caption{Bar chart for process metrics importance from the trained XGB model.}
    \label{fig:xgboostImportance-process}
\end{figure}
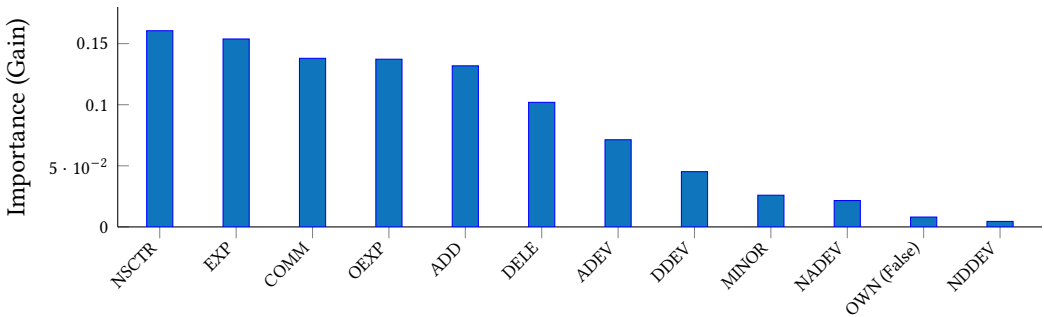

\paragraph{XGB Results.} 
XGB measures the importance of metrics in terms of Information Gain (IG). Regarding \textbf{process metrics}, top-ranked \textbf{NSCTR} (16.05\%) and \textbf{EXP} (15.35\%), with \textbf{COMM}, \textbf{OEXP}, and a second instance of \textbf{OEXP} scoring between $\sim$13–14\% (see Figure~\ref{fig:xgboostImportance-process}). These results are consistent with the RF-based findings, reaffirming that \textbf{developer experience} and the \textbf{breadth of file/package-level changes} are dominant predictors of RMCs. The repetition of OEXP highlights its consistent impact across learning models. Overall, these results strengthen the hypothesis that \textbf{developer history and cross-cutting modifications} are key explanatory factors in process-driven code evolution.

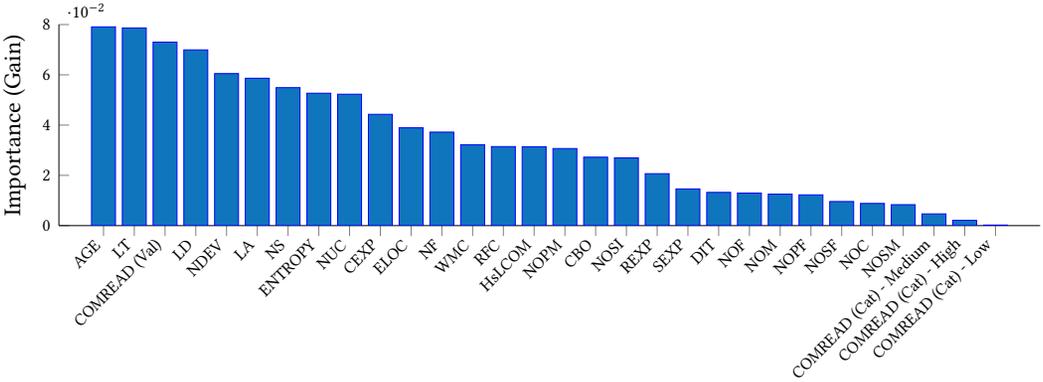
\begin{figure}[t]
    \centering
    \begin{adjustbox}{width=1\textwidth}
\begin{tikzpicture}
\begin{axis}[
    ybar,
    bar width=10pt,
    height=4.5cm,
    width=16cm,
    ymin=0,
    ymax=0.08,
    ylabel={Importance (Gain)},
    xtick=data,
    xticklabel style={
        rotate=45,
        anchor=east,
        font=\scriptsize
    },
    yticklabel style={
        font=\scriptsize
    },
    symbolic x coords={
        AGE, LT, COMREAD (Val), LD, NDEV, LA, NS, ENTROPY, NUC, CEXP,
        ELOC, NF, WMC, RFC, HsLCOM, NOPM, CBO, NOSI, REXP, SEXP,
        DIT, NOF, NOM, NOPF, NOSF, NOC, NOSM, COMREAD (Cat) - Medium,
        COMREAD (Cat) - High, COMREAD (Cat) - Low
    },
    axis lines*=left,
    enlarge x limits=0.05,
    point meta=y*100, 
    title={},
]

\addplot+[ybar, fill=acm-blue, draw=blue] coordinates {
    (AGE,                 0.0790894785)
    (LT,                  0.0786565584)
    (COMREAD (Val),         0.0730135517)
    (LD,                  0.0699207796)
    (NDEV,                0.0604970508)
    (LA,                  0.0586420070)
    (NS,                  0.0548968811)
    (ENTROPY,             0.0526816984)
    (NUC,                 0.0522910193)
    (CEXP,                0.0442870204)
    (ELOC,                0.0389282207)
    (NF,                  0.0372052889)
    (WMC,                 0.0322056296)
    (RFC,                 0.0314516181)
    (HsLCOM,              0.0314149054)
    (NOPM,                0.0306706840)
    (CBO,                 0.0272402811)
    (NOSI,                0.0269473281)
    (REXP,                0.0206719115)
    (SEXP,                0.0145958264)
    (DIT,                 0.0132247744)
    (NOF,                 0.0129572480)
    (NOM,                 0.0125288955)
    (NOPF,                0.0122431791)
    (NOSF,                0.0095846334)
    (NOC,                 0.0088264656)
    (NOSM,                0.0083349735)
    (COMREAD (Cat) - Medium,   0.0046705965)
    (COMREAD (Cat) - High,     0.0021005853)
    (COMREAD (Cat) - Low,      0.0002209096)
};

\end{axis}
\end{tikzpicture}
\end{adjustbox}
    \vspace{-2.2em}
    \caption{Bar chart for product metrics importance from the trained XGB model.}
    \label{fig:xgboostImportance-product}
\end{figure}

Regarding \textbf{product metrics},  \textbf{AGE} of a file (7.90\%) and its \textbf{length before modification (LT, 7.86\%)} emerged as the most informative (see Figure~\ref{fig:xgboostImportance-product}). These were followed by \textbf{COMREAD (Val)} (7.30\%), \textbf{LD} (6.99\%), and \textbf{NDEV} (6.05\%). These findings mirror the earlier MDG-based RF results, where code \textbf{maturity}, \textbf{comprehensibility}, and \textbf{developer interaction} were pivotal. The emphasis on \textbf{readability} and \textbf{historical change patterns} indicates that certain design and evolution characteristics make a file more prone to refactorings or maintenance.

\begin{figure}[t]
    \centering
    \begin{adjustbox}{width=1\textwidth}
\begin{tikzpicture}
\begin{axis}[
    ybar,
    bar width=7pt,
    height=4.5cm,
    width=18cm,
    ymin=0,
    ylabel={Importance (Gain)},
    xtick=data,
    xticklabel style={
        rotate=45,
        anchor=east,
        font=\scriptsize
    },
    yticklabel style={
        font=\scriptsize
    },
    symbolic x coords={
        COMREAD (Val), NSCTR, COMM, EXP, OEXP, AGE, ADD, NS, DELE, LT,
        LA, NDEV, CEXP, LD, NUC, ELOC, ENTROPY, NF, WMC, NOPM,
        RFC, NOSI, HsLCOM, CBO, DDEV, DIT, ADEV, SEXP, REXP, NOPF,
        NOF, NADEV, NOM, NOSF, NOC, OWN (True), NOSM, COMREAD (Cat) - High,
        MINOR, COMREAD (Cat) - Medium, NDDEV
    },
    axis lines*=left,
    enlarge x limits=0.05,
    scaled y ticks=base 10:2, 
    point meta=y*100,          
    title={},
]

\addplot+[ybar, fill=acm-blue, draw=blue] coordinates {
    (COMREAD (Val), 0.064489798)
    (NSCTR, 0.062499286)
    (COMM, 0.055176154)
    (EXP, 0.051755229)
    (OEXP, 0.051290104)
    (AGE, 0.047136077)
    (ADD, 0.045938682)
    (NS, 0.041311109)
    (DELE, 0.040753771)
    (LT, 0.040205648)
    (LA, 0.038114548)
    (NDEV, 0.037916381)
    (CEXP, 0.033289238)
    (LD, 0.031257901)
    (NUC, 0.029725522)
    (ELOC, 0.028149221)
    (ENTROPY, 0.022171983)
    (NF, 0.021269364)
    (WMC, 0.020038973)
    (NOPM, 0.019462249)
    (RFC, 0.019134518)
    (NOSI, 0.019100030)
    (HsLCOM, 0.017388911)
    (CBO, 0.016529507)
    (DDEV, 0.016382093)
    (DIT, 0.015907779)
    (ADEV, 0.014984479)
    (SEXP, 0.014420960)
    (REXP, 0.014222538)
    (NOPF, 0.011862542)
    (NOF, 0.010771847)
    (NADEV, 0.009575113)
    (NOM, 0.009053445)
    (NOSF, 0.005261278)
    (NOC, 0.004745945)
    (OWN (True), 0.004286137)
    (NOSM, 0.004075212)
    (COMREAD (Cat) - High, 0.003046790)
    (MINOR, 0.002627276)
    (COMREAD (Cat) - Medium, 0.002551441)
    (NDDEV, 0.002120922)
};

\end{axis}
\end{tikzpicture}
\end{adjustbox}
    \vspace{-2.2em}
    \caption{Bar chart for product and process metrics importance according to  XGBoost.}
    \label{fig:xgboostImportance}
\end{figure}
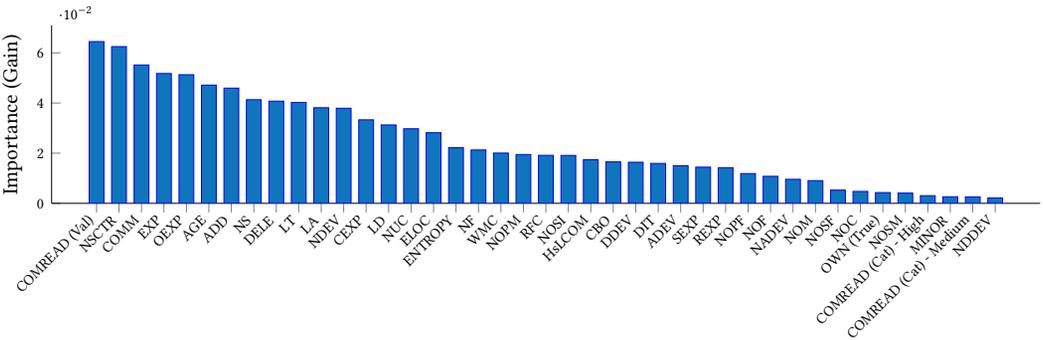

Finally, when combining all metrics in the XGB training (see Figure~\ref{fig:xgboostImportance}), \textbf{COMREAD} surfaced as the most important variable (6.45\%), affirming the \textbf{centrality of code comprehensibility} in predicting RMs. This was followed by \textbf{NSCTR} (6.25\%), the metric capturing change spread across packages, and \textbf{COMM} (5.52\%), \textbf{EXP} (5.18\%), and \textbf{OEXP} (5.13\%). The consistency across modeling techniques and metric types provides strong evidence that \textbf{code readability, developer experience, and change scope} are jointly the most critical signals for understanding why RMCs occur.

\begin{keyTakeAways}[Metrics Importance]
    Developer experience, code readability, and change scope are the strongest predictors of RMCs, with Random Forest emphasizing ownership and commits, and XGB highlighting readability and change propagation.
\end{keyTakeAways}

\subsubsection{Investigating Metrics Correlation with Refactoring Motivation Categories}
\label{sec:results-correlation}
The AD test results allowed us to reject the null hypothesis on normal data distribution ($H_{0\mathcal{N}}$),  therefore we must rely on non-parametric correlation analysis using \textbf{Spearman’s $\rho$} and \textbf{Kendall’s $\tau$}. 

\begin{figure}[t]
    \centering
    \includegraphics[width=\linewidth]{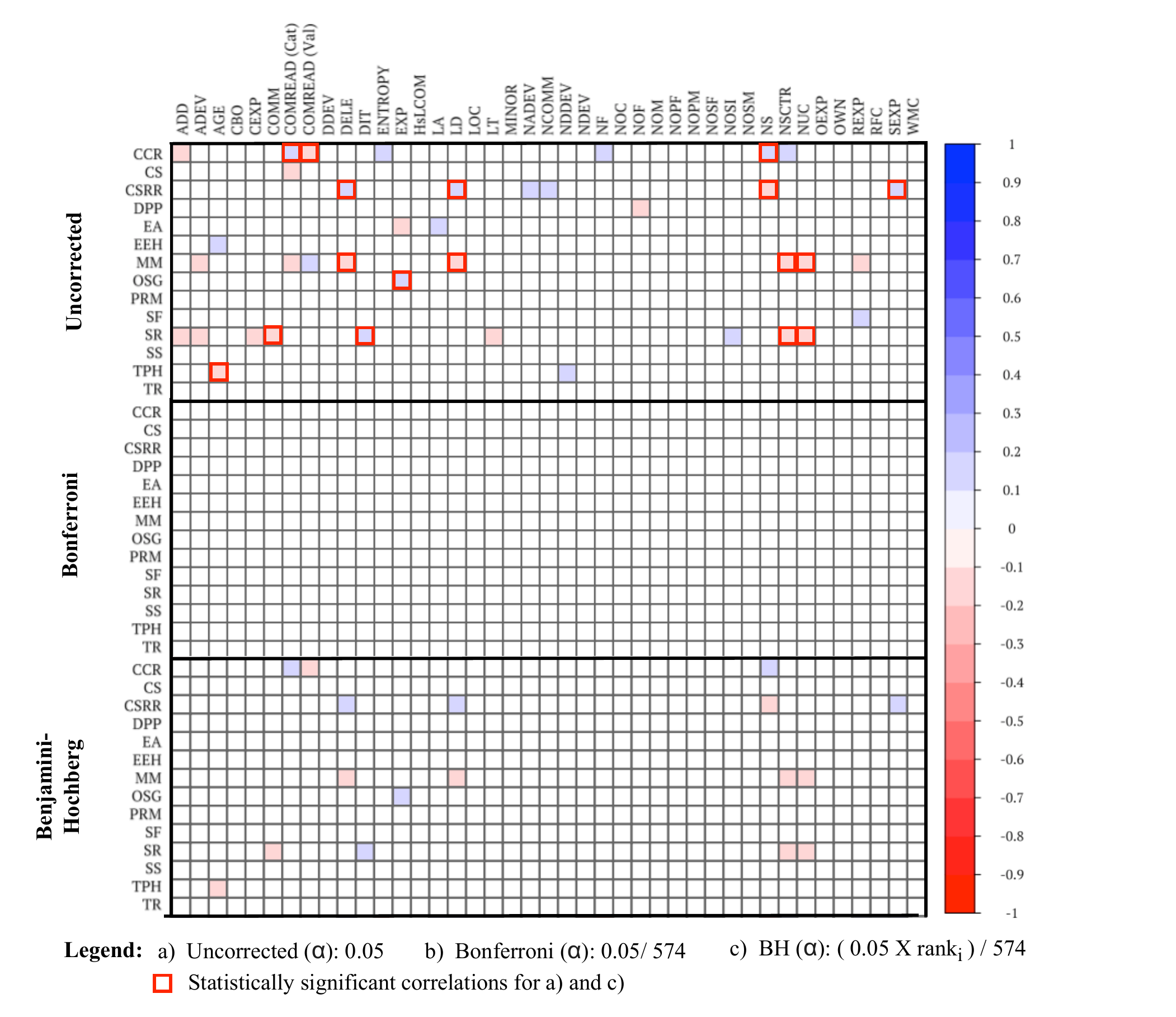}
    \caption{Heat-map representing the Spearman correlation matrix of RMCs with SMs for the uncorrected, Bonferroni corrected, and BH corrected probabilities.}
    \label{fig:correlation-spearman}
\end{figure}

\paragraph{Spearman Correlation Results.}
Figure~\ref{fig:correlation-spearman} displays the correlation matrix, both uncorrected and adjusted with Bonferroni and BH corrections. Out of 574 tests, we could reject the null hypothesis H$_{0}$ in 39 uncorrected correlations. After correction, we could only reject the null hypothesis H$_{0}$ in \textbf{17 cases} following the BH method.

\begin{figure}[t]
    \centering
    \includegraphics[width=\linewidth]{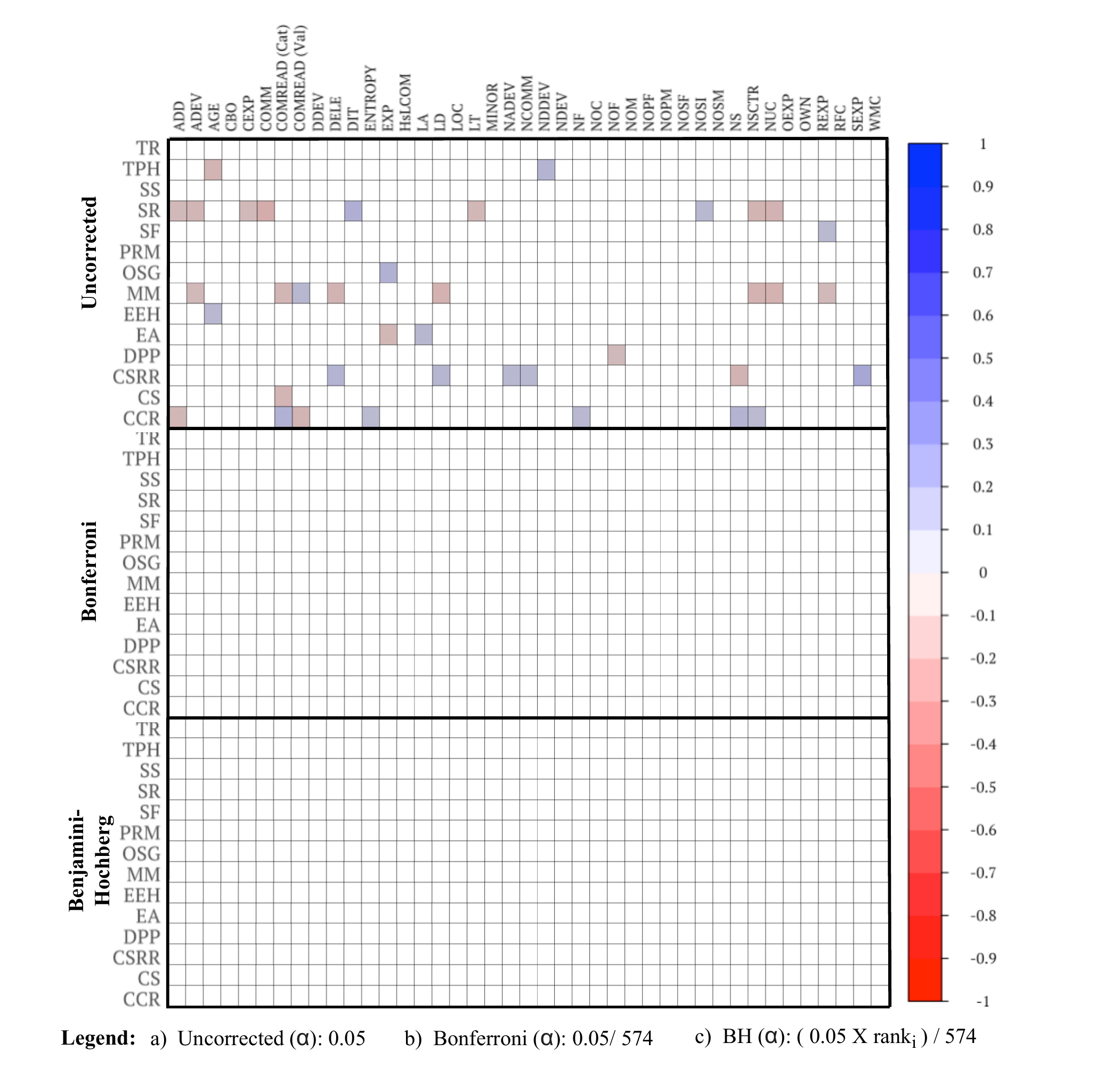}
    \caption{Heat-map representing the Kendall correlation matrix of RMCs with SMs for the uncorrected probabilities.}
    \label{fig:correlation-kendall}
\end{figure}

\paragraph{Kendall Correlation Results.}
The uncorrected Kendall $\tau$ allowed us to reject the null hypothesis H$_{0}$ in 39 out of 574 cases (see Figure~\ref{fig:correlation-kendall}), consistent with Spearman's. However, no significant results remained after applying Bonferroni or BH corrections.  

\paragraph{Metrics-based Results.}
The correlation analysis revealed several noteworthy, albeit weak, trends between specific metrics and RMs. Metrics like \textbf{COMREAD}, \textbf{NF}, \textbf{NSCTR}, and \textbf{ENTROPY} showed weak positive correlations with \textbf{code clarity (CCR)} motivations, suggesting that developers may prioritize readability and the modular spread of changes when aiming to clarify code. Interestingly, \textbf{COMREAD} and \textbf{NOF} were negatively correlated with \textbf{code standardization (CS)} motivations, possibly indicating that overly complex or large field declarations may hinder standardization efforts.

For \textbf{code simplification and removal of redundancy (CSRR)}, weak positive associations were found with \textbf{DELE}, \textbf{LD}, \textbf{NADEV}, and \textbf{SEXP}, reflecting a tendency to simplify code where there is significant churn and developer interaction. \textbf{LA} positively correlated with \textbf{abstraction-related RMs}, while \textbf{EXP} showed a weak negative relationship, possibly indicating that fewer experienced teams may abstract more aggressively.

The \textbf{AGE} metric was negatively linked to \textbf{technical performance and hygiene (TPH)} motivations, suggesting that older files might be neglected in performance-focused refactoring. For \textbf{structural reorganization (SR)}, \textbf{DIT} showed a weak positive correlation, highlighting deeper inheritance trees as a refactoring trigger, whereas \textbf{NUC} and \textbf{NSCTR} showed negative ones, implying that files with broader or frequent changes may be less targeted for structural reshaping.

Lastly, \textbf{REXP} correlated weakly positively with RMs tied to \textbf{supporting new functionalities (SF)}, while \textbf{COMREAD}, \textbf{ADEV}, and \textbf{REXP} showed weak negative correlations with \textbf{maintenance and modernization (MM)}, suggesting that highly readable or frequently changed code may not always be the primary target for modernization.

\begin{keyTakeAways}[Scarce Correlation and Statistical Significance]
    Out of 574 tests, Spearman’s $\rho$ identified 17 significant correlations after BH correction, revealing weak but interpretable links, e.g., code clarity (CCR) with readability and entropy, and maintainability (MM) with deletion and change frequency. Conversely, Kendall’s $\tau$ yielded no significant results post-correction. 
\end{keyTakeAways}

\paragraph{Answering RQ$_3$} Our findings revealed that both process and product metrics play complementary roles in predicting RMs. RF’s MDA favored contributor dynamics and modularity, highlighting metrics like ADEV, MINOR, and NS, while its MDG emphasized experience (EXP, OEXP) and code change intensity (COMM, NSCTR). Product-wise, structural complexity (NS, NUC) and readability (COMREAD.Val) emerged as important signals. XGB corroborated these trends: for process metrics, developer experience and file/package change spread (EXP, NSCTR) stood out; for product metrics, file age (AGE), size (LT), and readability again topped the list. Notably, across both models, developer engagement, code comprehensibility, and structural change scope consistently ranked as key predictors of RMCs. The findings underscore a dual influence of social and technical dimensions in shaping code evolution.

\begin{keyTakeAways}[Metrics and Refactoring Motivations]
Both RF and XGB models converge on three main predictors of RMs: \textbf{developer experience}, \textbf{code readability}, and \textbf{change scope}. RF highlights developer ownership and activity (e.g., EXP, OEXP, COMM), while XGB reinforces the role of code comprehensibility (COMREAD) and structural evolution (NSCTR, AGE). These findings collectively underscore the intertwined effect of human factors and code characteristics on driving refactoring decisions.

\end{keyTakeAways}

\section{Discussion}
\label{sec:discussion}

We leverage the following section to discuss the key findings of our study by relating the outcomes to every research question and identifying actionable implications for researchers and practitioners. We also address the applicability of our results in favour of refactoring recommendation systems.

As shown in RQ$_1$, LLMs agreed with human judgments in 80\% of the cases but only 47\% matched RMs from the reference study. Systematic disagreements emerged, as confirmed by Bowker’s test, with LLMs tending to interpret RMs based on local code cues, names, syntax, and immediate readability, rather than broader architectural goals. For instance, RMs for MM and MC saw substantial misalignment, with LLMs overlooking structural intentions behind these changes. In contrast, Pull Up operations had a near-perfect alignment, likely due to their localized semantics.

Human raters, instead, often based their evaluations on project-wide consistency, long-term maintainability, and design abstraction, factors that are invisible in the local code snapshot available to the LLM. This gap reveals that while LLMs can mimic reasoning for simple refactorings, they fall short in capturing RMs rooted in architecture or foresight.

\begin{implications}[Local vs Structural Reasoning]
LLM-generated RMs should be interpreted as partial approximations. They are effective for routine, localized changes but unreliable for strategic or architectural reasoning. To close this gap, future tools could expand LLM context with project-wide metadata or history-aware embeddings.
\end{implications}
According to RQ$_2$, only 53\% of LLM-generated RMs aligned exactly with the prior reference study. However, in 22\% of the cases, LLMs enriched the rationale by adding context-specific nuances or detailing effects on testability, naming conventions, or structural clarity. These elaborations were never contradictory; instead, they made tacit developer knowledge explicit.

This ability to extend RMs aligns with the finding that LLMs consistently emphasized clarity, maintainability, and readability. For example, when humans described motivation as “renaming,” the LLM often contextualized it in terms of onboarding ease or cognitive load, surfacing deeper reasoning that might otherwise be lost.

\begin{implications}[Enriching Motivation Clarity]
LLMs offer value by articulating implicit reasoning. Their output can help with onboarding, documentation, or code review, surfacing hidden RMs that developers rarely write down. This supports knowledge retention beyond the act of coding.
\end{implications}

The RQ$_3$ analysis of 385 samples revealed that over 55\% of all RMs were aimed at improving code clarity (CCR) or simplifying redundant logic (CSRR). Most RMs stemmed from pragmatic needs, naming clarity, method simplification, and parameter cleanup, rather than enforcing design principles or structural overhauls. Refactoring was thus largely incremental, not architectural.

This aligns with our earlier findings: LLMs were effective when refactorings were localized and incremental, and struggled when RMs depended on broader system-level reasoning. These trends match developer behavior observed in earlier work \cite{esposito2023uncovering}.

\begin{implications}[Pragmatism Over Principles]
Developers often refactor for short-term clarity, not ideal long-term architecture. Tools that surface refactoring suggestions should prioritize simplicity, understandability, and friction reduction over theoretical design goals.
\end{implications}
Our machine learning and correlation analysis inferences revealed that while some metrics (e.g., \texttt{EXP}, \texttt{OEXP}, \texttt{COMREAD}) were top-ranked by importance models (RF/XGB), their correlation with motivation categories was weak and often statistically insignificant. For example, only 17 correlations passed significance after correction in Spearman’s analysis, and none did in Kendall’s.

This suggests metrics alone lack the semantic richness to explain why developers refactor. However, their role as signals is non-negligible: metrics related to developer engagement, file age, and code readability consistently ranked high in feature importance scores, indicating they can guide or contextualize predictions.

\begin{implications}[Metrics in a Supporting Role]
Metrics do not reflect motivation directly but can inform LLMs of context and likelihood. They are better used as input features or filters in hybrid systems rather than as standalone predictors.
\end{implications}

Our research stemmed from the idea of supporting developers with a catalogue of SMs, informative on the behaviour of the drawn RMCs. A question that naturally arises from this research is whether our findings support the development of LLM-based refactoring recommendation systems. And the response is tentatively affirmative. LLMs showed a satisfactory ability to classify and justify RMs, and they were particularly resilient when justifying simple, localized refactoring intentions. They struggled with architectural or cross-cutting RMs and required human judgment to resolve open cases. The weak correlation between metrics and RMs means that a metric-only system would be unreliable. The only plausible direction is, therefore, hybrid systems: metric-informed LLMs that blend behavioral hints with natural language reasoning to suggest refactorings and explain why they do so.

\begin{implications}[LLM-Guided Refactoring Systems]
LLMs can be the basis of explanation- and clarity-focused refactoring recommendation systems. To make them more reliable, these systems can include SMs as well as project-level context, so that recommendations reflect both code attributes and probable developer intention.
\end{implications}

Our results highlight \textbf{key limitations and opportunities for research} in Generative AI-assisted software engineering. First, the systematic differences between LLM and human RMs suggest the need to explore methods for enriching LLM inputs with higher-order architectural or historical context. Researchers should investigate techniques such as long-context prompting, integration of architectural models, and retrieval-augmented generation (RAG) that can provide LLMs with broader system-level signals. The observed ability of LLMs to extend developer RMs also opens avenues for studying how to formalize tacit knowledge and how to balance verbosity with precision in generated justifications. Additionally, the weak correlation between metrics and motivation categories underscores a gap in our current abstraction tools. Research should re-evaluate which metrics meaningfully reflect cognitive developer behavior or develop new hybrid indicators that better represent intent. Finally, the potential of hybrid LLM-metric models invites empirical validation: how can we measure trust, usability, and impact of such systems in real-world settings?

\textbf{Practitioners} can leverage LLMs as assistants in tasks that require expressing, documenting, or reviewing code change rationales. Our results show that LLMs are especially useful in surfacing developer intent for routine, local refactorings and in articulating implicit reasoning that is often left undocumented. Teams can integrate LLMs into code review pipelines or commit templates to clarify RMs and make them traceable over time. However, caution is needed: for refactorings that touch architectural layers or involve non-local effects, LLMs should be treated as suggestive rather than authoritative. Moreover, while SMs remain useful, especially those reflecting experience and activity, they should complement, not substitute, LLM output. A practitioner-facing system that combines LLM insights with contextual metrics can support better onboarding, reduce misunderstanding in reviews, and help preserve architectural consistency by making rationale explicit and shareable.

\section{Threats to Validity}
\label{sec:Threat}
This section discusses the threats to validity, including internal validity, external validity, construct validity and reliability. 
Moreover, we explain the different adopted tactics~\citep{Wohlin2000}. 

\textbf{Construct Validity}. We acknowledge that using LLMs for \emph{semi-supervised multi-class} classification may pose a threat to the results of this study, as it assumes the artificial intelligence-based feedback to be correct. However, to minimize this threat, we considered using state-of-the-art as well as well-valued distilled LLMs such as Marco-o1 and Mistral NeMo, among others. Similarly, we provided the models with localized version control data related to the RM to support the prompt with project context, which might result in local reasoning output from the LLMs. We considered supporting the models with such level of local context to be the first approach for leveraging LLMs in RM identification, as no previous study had used them for such a task. Nevertheless, we intend to explore the potential of further prompting techniques, e.g., using multi-turn prompts and system-level metadata, to provide the model with a more holistic context level. Silva et al~\citep{silva2016we} had already considered the threat of using RMT for mining the commits in which the refactorings were performed as some classifications may be missed. However, 
we still consider using RMT to provide consistency to our results, as one of our goals is to validate the ground truth motivations provided by developers. Furthermore, RMT still remains the state-of-the-art refactoring mining tool to date. We also identify as a threat to the validity of our results the use of the refactoring commits, and their corresponding RMs, sourced from the reference study, which were derived from projects we did not mine with RMT in our study due to computational resource limitations. We tackled this threat by mining the SM related to the refactoring commit in which the affected RMs from the reference study were identified, and hence, we had all the required data to still consider these refactoring observations as ground-truth. 

\textbf{Internal Validity}. One of the main outcomes of this study is the publication of a set of metrics to guide developers in performing refactoring activities. These metrics are based on the human-based motivations as well as AI-based motivations obtained from the performed classification, hence we acknowledge the threat that the defined refactoring guideline metrics may not provide an accurate instruction to commit refactoring. Missing motivations from the developers and potential motivations yet to be discovered by analyzing further projects may contribute to the possible lack of accuracy. However, we understand that the motivations provided by human developers~\citep{silva2016we} provide the closest feedback to the real motivations for committing refactoring, and therefore, reduce this type of threat. We acknowledge the existing threat to validity in considering OSS LLMs for the motivation identification stage of the analysis. Leveraging the latest state-of-the-art models depends on budget and the specific required computational power. To tackle this threat, we searched for open-source alternatives that claim to present similar results as the current state-of-the-art reasoning models.

\textbf{External Validity.} This study only considers open source projects, based on the Java programming language and hosted by GitHub. Therefore, the results of this study, even though they aim to cover a wide range of projects, cannot claim to apply to systems outside the open-source community or projects developed in a different programming language. However, the presented study plan aimed to analyze all the existing refactoring activity of 124 projects and all their mineable refactoring types supported by the adopted mining tool, which, if the results provide a clear guideline of metrics, would provide one of the most consistent recommendation guidelines for refactoring activity in the field.  

\textbf{Reliability.} The planned data analysis is presented in a format that aims to provide answers to both the refactoring cases in which the motivation is identifiable and when discrepancies may exist and therefore already considered motivations may not fit for the refactoring case. Similarly, the choice of statistical tests detailed in this article aims at covering the possible statistical assumptions that data that is differently distributed may require for performing statistical testing. The source data provides a considerable set of GitHub-hosted projects. Yet, given the existing number of projects published on the mentioned platform, only a small portion of the samples was considered. Therefore, we assume as a threat the hypothetical differences in results if the same analysis were implemented in a different sample of projects. However, the validation of the developer-based motivations through the results of this study would minimize the presented threat as we understand that the motivations presented by developers are not the result of their performance in the analyzed projects but the result of their entire development career and hence the product of a larger number of projects.

\section{Related Work}
\label{sec:relworks}
This section reviews related work on refactoring recommendations, outlining the current state of research and presenting a detailed comparison with our approach in Table~\ref{tab:RelatedWork}.

Software refactoring can improve different aspects of software quality, such as readability and maintainability \citep{nyamawe2018recommending}. However, refactoring is also a time-consuming task and, as such, should be appropriately prioritized \citep{kurbatova2020recommendation}.  Therefore, to meet practitioners' needs, we must narrow the refactoring scope via recommendations \citep{alizadeh2018interactive}.
\citet{nyamawe2018recommending} investigated software refactoring, introducing a novel approach that evaluates refactoring solutions based on traceability and source code smell using entropy-based and traditional metrics, respectively. The authors found that their approach outperformed traditional metric-based recommendations in 71\% of the examined cases.
\citet{kaur2021brief} introduced a multi-objective optimization technique to generate refactoring solutions that maximize software quality, code smell severity, and consistency with class importance. They addressed the limitations of existing approaches, which focus solely on software enhancement and traditional software quality. Using a multi-objective spotted hyena optimizer (MOSHO), they evaluated refactoring solutions on five open-source datasets. The authors showed MOSHO's effectiveness in leveraging class importance and smell severity scores.

\citet{kurbatova2020recommendation} proposed a novel approach for recommending Move Method refactoring, mainly useful when methods rely heavily on external class members. Existing approaches, often heuristic-based, suffer from limitations such as metric dependency and manual threshold setting. On the other hand, the proposed approach leverages code2vec, a path-based code representation capturing syntactic and semantic information, to train a machine learning classifier. The empirical study on open-source projects and a synthetic dataset showed that the proposed approach outperformed existing tools like JDeodorant and JMove.

\citet{alizadeh2018interactive} proposed a dynamic, interactive refactoring recommendation approach to address the challenges of managing software complexity and facilitating software evolution. The proposed approach leverages NSGA-II~\citep{deb2002fast} to suggest refactorings while considering developer feedback, aiming to improve software quality while minimizing deviation from the initial design. They evaluated the approach on open-source and industrial projects, demonstrating its superiority over existing search-based techniques and fully automated tools.
Our work does not focus only on the mere refactoring effort but also on the developer's perspective on refactoring; more specifically, we investigate the motivations that lead developers to refactor specific portions of the code base.

More recently, \citet{jiang2025automated} proposed a novel approach to suggesting where local variable refactorings should be extracted and how the proposed suggestions should be extracted. Their presented alternative, and yet insightful tool \textit{ValExtractor+} on how to approach refactoring recommendation emphasized not only what local expressions should be extracted, but also how these expressions should be efficiently extracted. Their resulting recommendations were able to suggest correctly refactoring recommendations on where state-of-the-art tools were not successful, and their recommendations were approved and merged into the considered software projects.

A considerable step forward has been taken, proposed by ~\citet{liu2025exploring} exploring the potential use of LLMs in automated software refactoring. In their work, they explored real constraints such as (i) comparing the correctness of LLM-generated refactoring recommendations to that of human experts, and (ii) investigating the risks in terms of changes in the functionality of the code and syntax errors when introducing LLM-based refactoring within the codebase. Following the research on the application of LLMs in software refactoring, \citet{zhang2025move} proposed the \textit{MoveRec} tool focusing on refactoring recommendations for the Move Method refactoring type. For such a granularity level, they presented a deep learning and LLM information-based new refactoring recommendation system and compared it with existing state-of-the-art methods such as PathMove or JMove, among others. Their results demonstrated an improvement in effectiveness towards existing tools in terms of F1, ranging from 9.4\% to 53.4\%.
\begin{table}[hb]
\tiny
\renewcommand{\arraystretch}{1.3}
\centering
\caption{Related work on the existing previous research on Refactoring Recommendation.}
\label{tab:RelatedWork}
\begin{tabular}{p{1.8cm}|p{2.8cm}p{2.2cm}p{2.8cm}p{2.3cm}}\hline
	&	\citet{nyamawe2018recommending}	&	\citet{kurbatova2020recommendation}	&	\citet{alizadeh2018interactive}	&	\citet{kaur2021brief}	\\	\hline
\textbf{Granularity}	&	Method	&	Method	&	Class	&	Class, Method, Package	\\	
\textbf{PL}	&	Java	&	Java	&	Java	&	Java	\\	
\textbf{Projects}	&	1	&	14	&	10	&	5	\\	
\textbf{Analysis Scope}	&	Accuracy, Qualitative assessment	&	Quantitative assessment	&	Tool demonstration and evaluation	&	Tool demonstration and evaluation	\\	
\textbf{Dataset}	&	iTrust	&	JMove, synthetic dataset	&	-	&	-	\\	
\textbf{Model}	&	Own algorithm	&	SVM	&	NSGA-II-Innovization	&	MOSHO	\\	
\textbf{Refactoring  Rec.}	&	JDeodorant, JMove, ARIES	&	JDeodorant, JMove	&	Jdeodorant	&	NSGA-II, MOPSO, Mono-GA, \citet{kessentini2017detecting}, JDeodorant	\\	
\textbf{Mining tools}	&	-	&	-	&	-	&	-	\\	
\textbf{Refactorings}	&	2	&	1	&	11	&	25	\\	
\textbf{Discussion}	&	Yes	&	Yes	&	Yes	&	Yes	\\	
\textbf{Results}	&	No	&	No	&	Yes	&	No	\\	
\textbf{Scripts}	&	No	&	No	&	Yes	&	No	\\	
\textbf{Datasets}	&	Yes	&	Yes	&	Yes	&	No	\\	\hline
	&	\citet{pantiuchina2020developers}	&	\citet{liu2025exploring}	&	\citet{jiang2025automated}	&	\citet{zhang2025move}	\\	\hline
\textbf{Granularity}	&	Pull Request	&	Source Code	&	Local Variable	&	Method	\\	
PL	&	Java	&	Java	&	Java	&	Java	\\	
\textbf{Projects}	&	150	&	20	&	119	&	58	\\	
\textbf{Analysis Scope}	&	Accuracy, Qualitative assessment, Taxonomy 	&	Accuracy	&	Accuracy, Qualitative assessment	&	Accuracy	\\	
\textbf{Dataset}	&	Criteria-based GitHub mining	&	\citet{grund2021codeshovel}	&	Defects4J, GrowingBugs	&	-	\\	
\textbf{Model}	&	Mixed effect generalized linear model	&	ChatGPT, Gemini	&	OpportunityAdvisor, solutionAdvisor	&	MoveRec	\\	
\textbf{Refactoring  Rec.}	&	-	&	-	&	JRRT	&	PathMove, JDeodorant, JMove, RMove	\\	
\textbf{Mining tools}	&	RefactoringMiner	&	-	&	RefactoringMiner	&	RefactoringMiner	\\	
\textbf{Refactorings}	&	14	&	9	&	1	&	1	\\	
\textbf{Discussion}	&	Yes	&	Yes	&	Yes	&	Yes	\\	
\textbf{Results}	&	Yes	&	Yes	&	Yes	&	No	\\	
\textbf{Scripts}	&	Yes	&	Yes	&	Yes	&	No	\\	
\textbf{Datasets}	&	Yes	&	Yes	&	Yes	&	No	\\	\hline
	&	\citet{pomian2024next}	&	\textbf{Our work}	&		&		\\	\hline
\textbf{Granularity}	&	Method	&	Source Code	&		&		\\	
\textbf{PL}	&	Java	&	Java	&		&		\\	
\textbf{Projects}	&	17	&	114	&		&		\\	
\textbf{Analysis Scope}	&	Accuracy, Survey	&	Descriptive analysis, Qualitative assessment, Taxonomy	&		&		\\	
\textbf{Dataset}	&	Community Corpus, Extended Corpus	&	~\citet{silva2016we}	&		&		\\	
\textbf{Model}	&	EM-Assist	&	Marco-o1, Deepseek R1, Mistral NeMo, Phi-4	&		&		\\	
\textbf{Refactoring}  Rec.	&	REMS, GEMS, Jdeodorant, Jextract, SEMI, LiveREF	&	-	&		&		\\	
\textbf{Mining tools}	&	RefactoringMiner	&	RefactoringMiner	&		&		\\	
\textbf{Refactorings}	&	1	&	103	&		&		\\	
\textbf{Discussion}	&	Yes	&	Yes	&		&		\\	
\textbf{Results}	&	Yes	&	Yes	&		&		\\	
\textbf{Scripts}	&	Yes	&	Yes	&		&		\\	
\textbf{Datasets}	&	Yes	&	Yes	&		&		\\	\hline
\end{tabular}%
\end{table}

Focusing on the insights hidden behind developers' refactoring motivations, \citet{pantiuchina2020developers} performed a large-scale study to understand why developers refactor code in open-source projects, complementing previous findings from developer surveys. They analysed 287,813 refactoring operations across 150 systems, examining the relationship between refactoring operations and process/product metrics.  The authors highlighted that a relationship exists between metrics and refactoring operations, along with a detailed taxonomy of refactoring motivations. 
Our study differs from Pantiuchina et al.'s~\citep{pantiuchina2020developers} approach in study design and goals. While ~\citep{pantiuchina2020developers} relied on keyword lookup and the RMT tool for PR analysis, we propose a novel methodology employing Language Model analysis, and therefore following the latest research trends on refactoring analysis. Moreover, while ~\citep{pantiuchina2020developers} focused on product quality and developer-related metrics, our study prioritizes established product and process metrics. Lastly, while the goal was to correlate product quality and developer metrics with refactoring in PRs, we aim to comprehensively analyze motivations for refactoring across various contexts and define a catalogue of motivations supported by a broader selection of metrics from established guidelines.

\section{Conclusion}
\label{sec:Conclusions}

Our study investigated the motivations behind performing code refactoring by employing LLMs on the version-control history data associated with the refactoring, compared their output to human rationale, as well as to the referenced prior study~\citep{silva2016we}. Moreover, we studied the extent to which product and process metrics can describe such RMs by analyzing their feature importance and correlation levels. Our results reveal that while LLMs agree with human judgment in most cases, they often diverge in deeper, architectural reasoning, favouring local, surface-level cues over systemic design intent. Still, LLMs demonstrate value in extending and clarifying RMs, surfacing implicit knowledge that is often under-documented.

We further showed that most LLM-derived RMs are pragmatic, driven by clarity and simplification, rather than architectural ideals. SMs to depict developer experience and code readability proved informative but insufficient as standalone signals, reinforcing the need for hybrid solutions.

Future research should focus on developing context-aware, LLM-guided refactoring recommendation systems that integrate both behavioural signals, such as RMs, and project-level insights. Similarly, future work should exploit the potential of further exploring prompting techniques, e.g., using multi-turn prompts, system-level metadata, or dependency graphs as prompt context. We believe such studies would provide LLMs with a more holistic analysis of the context involving the refactoring operation. LLMs can enhance traceability, documentation, and onboarding when paired with system context and developer interaction data. For researchers, this opens a space to explore enriched prompting, system-aware modeling, and empirical evaluation of hybrid systems. For practitioners, LLMs offer an opportunity to make the “why” behind code changes visible, fostering better collaboration and maintainability. LLMs do not just replicate developer intent; they can help understand, explain, and extend it. The next challenge is building tools that channel this capability with precision, context, and reliability.

{
\section*{Declarations}
\small
\textbf{Author Contributions:} We specify our contributions according to the CRediT taxonomy:

\begin{itemize}
    \item \textbf{Mikel Robredo:} Writing – Original Draft Preparation, Data Curation, Software, Data Curation. Mikel provided essential data preparation and contributed to drafting specific sections and contributed to writing the registered report.
    \item \textbf{Matteo Esposito:} Methodology, Writing – Original Draft Preparation. Matteo developed the primary research methodology and contributed significantly to writing the initial manuscript draft and registered report.
    \item \textbf{Fabio Palomba} Methodology, Validation, Writing – Review \& Editing, Supervision. Fabio critically validated the study’s results, supervised the research’s methodological approach, and reviewed the manuscript for academic rigor.
    \item \textbf{Rafael Peñaloza:} Formal Analysis, Supervision. Rafael performed a thorough formal analysis of data and statistical techniques and guided the research through supervision.
    \item \textbf{Valentina Lenarduzzi:} Conceptualization, Methodology, Supervision, Validation, Writing – Review \& Editing. Valentina led the conceptualization and methodological framing of the research, supervised the entire project, validated findings, and contributed to reviewing and refining the manuscript.
\end{itemize}
\textbf{Equal Author Contribution:} Mikel Robredo and Matteo Esposito contributed equally to this work.

\noindent \textbf{Conflict of Interest:}
We declare that we have no competing interests.

\noindent \textbf{Data Availability Statement:}
\label{sec:Replicability}
We provide our raw data and analysis results in our replication package hosted on Zenodo.\footnote{\url{https://doi.org/10.5281/zenodo.15209154}}

\noindent \textbf{Ethical Approval:}
Our work did not need ethical approval.

\noindent \textbf{Funding:}
This work has been funded by FAST, the Finnish Software Engineering Doctoral Research Network, funded by the Ministry of Education and Culture, Finland.

}

\bibliographystyle{ACM-Reference-Format} 
\bibliography{main}

\newpage
\appendix
\section*{Appendix}

In this appendix, we describe additional content to help the reader better understand the insights of the descriptions provided in the main text of this study.

\section{Supplementary material for the data collection}
\label{sec:AppendixA}

\subsection{Expanded description on the refactoring collection using RMT}
\label{sec:AppendixA_1}

RMT remains to be the state-of-the-art refactoring mining tool as it holds a refactoring detection accuracy close to 100\%~\footnote{\url{https://github.com/tsantalis/RefactoringMiner}}, which overcomes the capabilities of other existing tools such as \textsc{RefDiff}~\citep{silva2020refdiff}). RMT can detect 103 refactoring types through the analysis of how the Abstract Syntax Tree of a JAVA class/method has changed concerning one of the previous commits. Table~\ref{tab:refactorings} shows the categorization of the 103 refactorings of our study according to their types as defined by Fowler~\citep{fowler1999refactoring}. Out of the RTs that can be detected, only a subset of the detected RTs is classified in the original Fowler catalogue. Nonetheless, the RMT can also identify types that cannot be mapped to the original catalogue, such as the composite refactoring \textit{Move and Inline Method}.

We conducted the mining process with RMT on a workstation equipped with an Intel Core i9-13900KF CPU (24 cores, 32 threads), 64GB of RAM, and 7.27TB of disk storage. Similarly, we configured RMT to use a maximum share of RAM of 57GB, allowing other processes to continue running smoothly. Leveraging the custom options RMT offers, we launched the tool for each project to report the refactoring activity as a JSON file comprising each of the analyzed commits, the set of applied refactoring operations, as well as the classes/methods subject to them. Yet, projects \textit{liferay/liferay-plugins}, \textit{checksytle/checkstyle}, \textit{deeplearning4j/deeplearning4j}, and \textit{jOOQ/jOOQ} resulted in exceeding \textit{heap-space} error, which we could not resolve given our available computational resources. In addition, projects \textit{jersey/jersey}, \textit{crate/crate}, \textit{JetBrains/MPS}, \textit{siacs/Conversations}, \textit{kuujo/copycat}, \textit{bitfireAT/davroid} resulted in untraceable stalled processes. Consequently, after a prolonged period of multiple attempts, we opted to proceed without the mentioned projects. The discarded projects accounted for a total of 13 refactoring commits present in the reference study. However, since we could collect the SMs related to affected refactoring commits, and the replication package from the reference study provided the affected RT, we still considered the affected commits, and therefore their ground-truth refactoring motivations, in the study. We acknowledge the impact of this decision as a threat to validity in Section~\ref{sec:Threat}.

\subsection{Expanded description on the undergone sampling strategy}
\label{sec:AppendixA_2}

\paragraph{\textbf{Phase 1}:} Initially, we aimed to achieve enough representativeness for each of the mined projects and all the RTs with the drawn sample. For such a goal, we decided on setting a data structure to record the sampled refactoring observations, until all the projects had a minimum total of \textbf{3 refactorings}. Similarly, all the refactoring types had a minimum representation of the same number of refactorings on a global level.

\paragraph{\textbf{Phase 2}:} The previous phase did not reach the full representativeness of the mined 114 projects, as some projects did not detect refactorings from the RTs left to be sampled. Therefore, we expanded the sampling logic implemented in the previous phase and identified which projects remained under-sampled. We adopted the \textbf{reservoir sampling} technique to randomly pick the remaining refactorings~\citep{vitter1985random} to fill the under-sampled projects. Given that we knew the total number \textit{n} of refactorings for each of the projects remaining unrepresented, and we also knew the \textit{k} amount of sampled refactoring needed in each of the projects, reservoir sampling ensured that every refactoring observation within the set of mined refactorings per project had the same probability of being included on the set or \textit{reservoir} of refactorings still pending to be sampled per project. %

\paragraph{\textbf{Phase 3}:} Phases I and II resulted in a randomly chosen, yet balanced sample of 342 refactoring observations. Therefore, to ensure the target sample size to provide statistically significant results from the data analysis, we pooled all the mined refactorings together and performed pure random sampling with the remaining 43 refactoring observations.

\section{Supplementary material for the data analysis}
\label{sec:AppendixB}

\subsection{Selected LLM models}
\label{sec:AppendixB_1}

To balance performance and efficiency, we selected high-performing \textit{distilled} LLMs~\citep{xu2024survey}. Model distillation refers to the process of transferring knowledge from an LLM to a smaller one, aiming to preserve performance while reducing model complexity and computational cost (see Section~\ref{sec:Threat}). It is a complex task to know which model configuration leads to the most efficient answers when performing prompt engineering~\citep{son2025optimizing}. For this reason, we fixed the model temperature at 0.8, which provides a balance between creativity and coherence~\citep{davis2024temperature}.

\begin{itemize}
    \item \textbf{Marco-o1}~\citep{zhao2024marco}: A 7.6B parameter distilled version of OpenAI's o-1 model~\citep{zhong2024evaluation}, fine-tuned on curated CoT datasets. It uses Monte Carlo Tree Search (MCTS) and softmax-based scoring to explore reasoning paths, and shows strong performance across math, coding, and logic tasks.\footnote{\url{https://huggingface.co/AIDC-AI/Marco-o1}} The model configuration when performing the prompting included a context length of 4096 tokens.

    \item \textbf{Mistral NeMo}~\citep{MistralAI_2024}: A 12B model distilled from Mistral 7B~\citep{MistralAI_2023}, optimized for multi-turn reasoning, code generation, and instruction following. It benefits from alignment fine-tuning and supports 128K token inputs.\footnote{\url{https://huggingface.co/QuantFactory/Mistral-Nemo-Base-2407-GGUF} The model configuration when performing the prompting included a context length of 4096 tokens.}

    \item \textbf{DeepSeek R1}: A 14B distilled model from the DeepSeek-V3-Base series, trained with CoT examples and refined via human feedback and reinforcement learning. It matches o-1-level performance on reasoning benchmarks.\footnote{\url{https://deepseek.ai}} Given the bigger size of DeepSeek R1, we configured the context length of this model to be double the one from the previously adopted distilled models, i.e., 8129 tokens.

    \item \textbf{Microsoft Phi-4}~\citep{abdin2024phi}: A 14B model with a 16K token context window, trained on high-quality academic and technical data. It consistently outperforms larger models like GPT-4o in formal reasoning tasks.\footnote{\url{https://huggingface.co/microsoft/phi-4}} Similarly, as with DeepSeek R1, given that the size of Phi-4 was the same, we configured the context length limit to be the same, fixed at 8129 tokens.
\end{itemize}

To run the selected models locally, we used LM Studio\footnote{\url{https://lmstudio.ai}}, a desktop application designed for experimenting with LLMs in an offline environment. LM Studio integrates directly with the \textit{Hugging Face} model hub\footnote{\url{https://huggingface.co/models}}, allowing us to download and launch a variety of models easily. It also sets up a local server that mimics the OpenAI API interface~\footnote{\url{https://platform.openai.com/docs/api-reference/introduction}}, enabling smooth compatibility with tools built around OpenAI’s endpoints. For interaction and prompt execution, we relied on the Python \texttt{openai} library developed by~\citet{pypiOpenai}. All experiments were conducted on a high-performance workstation equipped with an NVIDIA GeForce RTX 4090 GPU.

\subsection{On the adopted CoT prompting strategy}
\label{sec:AppendixB_2}

\begin{figure}[t]
    \centering
    \includegraphics[width=\textwidth]{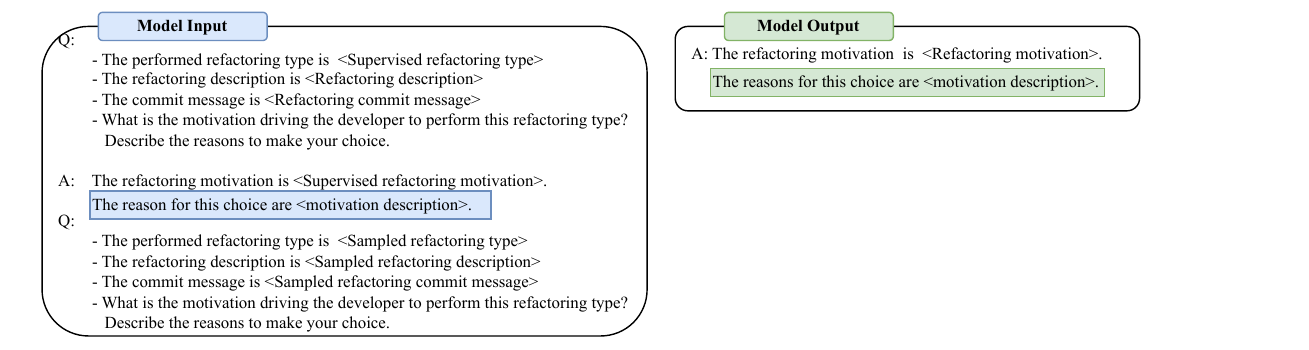}
    \caption{Example of Zero-Shot CoT prompt based on~\citet{kojima2022large}.}
    \label{fig:cot-learning}
\end{figure}

\paragraph{\textbf{CoT prompting}, } is a method that enhances the ability of LLMs to perform complex reasoning by eliciting a series of intermediate reasoning steps~\citep{wei2022chain}. CoT involves generating a "chain of thought", that is, a coherent sequence of natural language reasoning steps that lead to the final answer for a problem. It enables models to decompose multi-step problems, suggesting how they arrive at a particular answer. The process mimics a step-by-step thought process, similar to how humans solve complicated reasoning tasks. There are two primary ways to execute: Few-shot CoT prompting~\citep{wei2022chain} and Zero-shot CoT prompting~\citep{kojima2022large}. In our work, we adopted the latter since we did not want to introduce bias in the LLMs' responses based on the provided sample answers, and the ground-truth motivations extracted from the reference work did not include the entirety of the studied RTs. Figure~\ref{fig:cot-learning} provides an example pseudo-prompt following the Zero-shot CoT prompt strategy. Based on the required outputs for each of the defined RQs, we fine-tuned system and user prompt messages correspondingly, and made them available in the replication package (see Data Availability Statement). The last format of the shared prompt templates, and therefore the prompt messages used in the study, are the result of a progressive fine-tuning process where the quality of the LLM responses was discussed among the authors until a unanimous agreement was found.

\paragraph{\textbf{Structured JSON response}.} The expected model output consisted of a structured JSON response containing: (i) a brief answer/label with the requested answer, (ii) a concise \texttt{answer description}, and (iii) the underlying \texttt{reasoning} process involved in building the answer. This process was modified for each task prompting stage, correspondingly, based on each RQ. We instructed the LLMs to present their response in the defined JSON output. For that, we leveraged OpenAI's REST API~\footnote{\url{https://platform.openai.com/docs/guides/structured-outputs}} to enforce the prompted model to process the response output in JSON format.

\subsection{On the undergone Open-Coding.}
\label{sec:AppendixB_3}

Since we already had the validated RMs, this time we tasked models V1, V2, and V3 to perform open coding on the 385 RMs. Each model received
the detected RM, a description of the RM (generated by the best-performing model from a previous task), and an initially empty list of categories (called the \textbf{pooled list}) that could grow during the process. It is important to note that, following the design of an open coding process, the pool of motivation categories would start empty and greedily increase as more refactoring motivation observations are processed by the validation assistant models.

Each model was instructed to assign a clear motivation category (RMC) to the RM and explain why. The models were encouraged to reuse existing categories when possible, but they could also introduce new ones if needed. After V1 and V2 completed their coding, V3 reviewed the same RM and the two proposed categories. V3 then chose the final category based on three options:

\begin{enumerate}
    \item If V1 and V2 agreed, V3 accepted that shared category.
    \item If they disagreed, V3 picked one of the two based on their explanations.
    \item If neither fit well, V3 could suggest a new category, with justification.
\end{enumerate}

Likewise, given the absence of existing refactoring motivation categories and to respect the nature of an open coding task, we followed the zero-shot CoT learning strategy. Each of the models was instructed to consider the already existing categories provided in the user message. Yet, they were allowed to suggest a new motivation category if the existing categories in the list above did not fit the nature of the concerning motivation.

Finally, the human experts focused on a statistically representative sample (95\% confidence level, 5\% margin of error)~\citep{sandelowski1995sample}. The experts also repeated the open-coding process on the ground-truth dataset\cite{silva2016we} to compare and uncover additional developer motivations.

\section{List of Abbreviations from the detected Refactoring Types.}

\setcounter{table}{0}
\renewcommand{\thetable}{\Alph{section}.\arabic{table}}
\label{sec:AppendixC}
\begin{table}[h]
\centering
\caption{List of abbreviations for the Refactoring Types analyzed in the study.}
\label{tab:Abbreviations}
\tiny
\resizebox{\linewidth}{!}{%
\begin{tabular}{@{}p{0.27\linewidth}p{0.08\linewidth}|p{0.27\linewidth}p{0.08\linewidth}|p{0.27\linewidth}p{0.08\linewidth}@{}}
\hline
\textbf{Refactoring Type}       & \textbf{Abbr.} & \textbf{Refactoring Type}        & \textbf{Abbr.} & \textbf{Refactoring Type}       & \textbf{Abbr.} \\ \hline
Extract Variable                & EV               & Change Attribute Access Modifier & CAAM             & Remove Attribute Annotation     & RAA              \\
Change Parameter Type           & CPT              & Extract And Move Method          & EAMM             & Add Thrown Exception Type       & ATET             \\
Rename Parameter                & RenP             & Remove Attribute Modifier        & RAM              & Encapsulate Attribute           & EnA              \\
Extract Method                  & EM               & Change Return Type               & CRT              & Add Variable Modifier           & AVM              \\
Change Variable Type            & CVT              & Extract Interface                & EI               & Move And Inline Method          & MAIM             \\
Add Parameter                   & AP               & Rename Variable                  & RV               & Inline Method                   & IM               \\
Add Class Annotation            & ACA              & Modify Parameter Annotation      & MPA              & Reorder Parameter               & RParam           \\
Remove Method Annotation        & RMA              & Modify Attribute Annotation      & MAA              & Inline Attribute                & IA               \\
Rename Attribute                & RA               & Add Attribute Annotation         & AAA              & Rename Class                    & RC               \\
Add Parameter Annotation        & APA              & Parameterize Variable            & PV               & Parameterize Attribute          & PA               \\
Inline Variable                 & IV               & Modify Method Annotation         & MMA              & Localize Parameter              & LP               \\
Merge Parameter                 & MParam           & Remove Class Annotation          & RCA              & Remove Parameter Annotation     & RPA              \\
Merge Attribute                 & MerA             & Change Class Access Modifier     & CCAM             & Remove Variable Annotation      & RVA              \\
Add Method Annotation           & AMA              & Change Method Access Modifier    & CMAM             & Change Thrown Exception Type    & CTET             \\
Move Method                     & MM               & Add Class Modifier               & ACM              & Pull Up Method                  & PUM              \\
Move Attribute                  & MA               & Replace Loop With Pipeline       & RLWP             & Move Source Folder              & MSF              \\
Add Attribute Modifier          & AAM              & Move Class                       & MovC             & Extract Superclass              & ESup             \\
Extract Class                   & EC               & Modify Class Annotation          & MCA              & Replace Attribute With Variable & RAWV             \\
Replace Generic With Diamond    & RGWD             & Merge Conditional                & MCon             & Merge Class                     & MerC             \\
Invert Condition                & IC               & Add Method Modifier              & AMM              & Move And Rename Class           & MARM             \\
Replace Variable With Attribute & RVWA             & Replace Anonymous With Lambda    & RAWL             & Remove Variable Modifier        & RVM              \\
Rename Method                   & RM               & Remove Thrown Exception Type     & RTET             & Move Package                    & MP               \\
Move And Rename Attribute       & MARA             & Remove Class Modifier            & RCM              & Rename Package                  & RPack            \\
Remove Parameter                & RemP             & Remove Method Modifier           & RMM              & Split Parameter                 & SP               \\
Extract Attribute               & ExA              & Move And Rename Method           & MARM             & Split Attribute                 & SA               \\
Change Attribute Type           & CAT              & Move Code                        & MCode            & Pull Up Attribute               & PUA              \\
Push Down Method                 & PDM              & Split Method                     & SM               & Merge Variable                  & MV               \\
Split Package                   & SP               & Split Conditional                & SC               & Replace Attribute               & RA               \\
Merge Package                   & MPack            & Split Variable                   & SV               & Replace Anonymous With Class    & RAWC             \\
Merge Method                    & MerM             & Split Class                      & SClass           & Replace Pipeline With Loop      & RPWL             \\
Change Type Declaration Kind    & CTDK             & Replace Conditional With Ternary & RCWT             & Try With Resources              & TWR              \\
Remove Parameter Modifier       & RPM              & Push Down Attribute              & PDA              & Assert Throws                   & AT               \\
Add Parameter Modifier          & APM              & Add Variable Annotation          & AVA              & Merge Catch                     & MCat             \\
Collapse Hierarchy              & CH               & Extract Subclass                 & ESub             & Modify Variable Annotation      & MVA              \\
Parameterize Test               & PT               &                                  &                  &                                 &                  \\ \hline
\multicolumn{3}{l}{ (\textbf{Abbr.}: Abbreviation)}
\end{tabular}%
}
\end{table}

\section{List of Refactoring Motivation Categories per motivation family, along with their duplicate frequencies.}

\setcounter{table}{0}
\renewcommand{\thetable}{\Alph{section}.\arabic{table}}
\label{sec:AppendixD}
\begin{table}[t]
\centering
\caption{List of unique motivation categories and their respective frequencies grouped into the defined RMCs.}
\label{tab:motivation-categories-part1}
\tiny
\resizebox{\linewidth}{!}{%
\begin{tabular}{@{}l|l@{}}
\hline
\textbf{Motivation Category} & \textbf{Unique motivations (\# Occurrences)} \\
\hline
\textbf{\begin{tabular}[c]{@{}l@{}}Code Clarity\\and Readability\end{tabular}} & \begin{tabular}[c]{@{}l@{}}{Enhance} (2) / {Improve} (2) \textbf{Code Clarity} (11) {and Simplification} (12) / {and Maintainability} (2) / {and Abstraction} (11),\\ \textbf{Improve} {Readability} (3) / {Data Structure Semantics} (2), {Enhance} (12) / {Improve} (1) \textbf{Code} \textbf{Readability and Maintainability} (10),\\ {Enhance} (3) \textbf{Code Readability} (1) {and Consistency} (4) / {and Simplification} (1) / {by Isolating Logic} (1), {Enhance} (3) \textbf{Code Organization} (6) \\ {and Promote Reusability} (1), \textbf{Annotation Update for Deprecation} (2), \textbf{Package Organization Alignment} (3), \\ \textbf{Ensure Correctness and Consistency} (2), \textbf{Clarify} {Nullability Semantics} (3) / {Class Purpose} (1), \textbf{Enhance Readability} (6) \\ {and Maintainability} (11), \textbf{Improve Clarity by Isolating Complex Expressions} (1), \textbf{Enhance Code Conciseness and Readability} (1)\end{tabular} \\
\hline
\textbf{\begin{tabular}[c]{@{}l@{}}Maintainability\\and Modularity\end{tabular}} & \begin{tabular}[c]{@{}l@{}}\textbf{Enhance} {Modularity} (1) / {and Reusability} (14) / {Code Flexibility and Maintainability} (1) / {Code Modularity} (1), \textbf{Promote Reuse} (3),\\ \textbf{Improve Code Organization} (1), \textbf{Increase Flexibility} (7), {Improve} (1) \textbf{Code Quality} (1) {Improvement} (1),\\ \textbf{Centralize Shared Behavior} (1), \textbf{Reorganize Code Structure} (2), \textbf{Leverage Existing Implementation} (1)\end{tabular} \\
\hline
\textbf{\begin{tabular}[c]{@{}l@{}}Encapsulation\\and Abstraction\end{tabular}} & \begin{tabular}[c]{@{}l@{}}{Increase} (1) / {Improve} (1) / {Enhance} (2) \textbf{Encapsulation Breach} (1) / {Enhancement} (15), \textbf{Attribute Coupling Reduction} (1),\\ \textbf{Encapsulate Attribute} (1), \textbf{Reduce} {Global State and Enhance Encapsulation} (1) / {Class Coupling and Enhance Encapsulation} (1)\end{tabular} \\
\hline
\textbf{\begin{tabular}[c]{@{}l@{}}Testing\\and Reliability\end{tabular}} & \begin{tabular}[c]{@{}l@{}}\textbf{Reusability of Test Methods} (2), \textbf{Enhance Test Organization} (2), \textbf{Test Isolation Improvement} (2),\\ \textbf{Reduce Coupling and Enhance Testability} (2), \textbf{Enable} {Test Flexibility} (3) / {Test Execution} (1),\\ \textbf{Move Method to Facilitate Test Setup} (1), \textbf{Enhance Test Reliability} (1), \textbf{Adopt Modern Testing Framework} (2),\\ \textbf{Eliminate Unnecessary Variables} (3)\end{tabular} \\
\hline
\textbf{\begin{tabular}[c]{@{}l@{}}Code Simplification\\and Redundancy\\ Reduction\end{tabular}} & \begin{tabular}[c]{@{}l@{}}\textbf{Reduce Redundancy} (9), \textbf{Simplify Code} (4) {by Reducing Complexity} (4) / {Structure} (1) / {by Removing Unnecessary Annotations} (1) /\\ {by Removing Unnecessary Elements} (1) / {Simplify Code by Reducing Redundancy} (1), \textbf{Eliminate Redundant Code} (1),\\ \textbf{Remove} {Unnecessary Code Elements} (4) / {Redundant Parameters} (1), {Redundant Annotation} (1) / {Unnecessary Variable} (1),\\ \textbf{Simplify Parameter Handling by Reducing Argument Count} (1), \textbf{Reduce Code Duplication} (7) {and Enhance Flexibility} (1),\\ \textbf{Simplify Constructor Logic} (1), \textbf{Reduce Complexity} (16) {by Removing Unnecessary Code Segments} (2) /\\ {by Eliminating Unnecessary Abstractions} (1), \textbf{Eliminate Misleading or Unnecessary Information} (1),\\ \textbf{Simplify Constant Management} (1), \textbf{Eliminate Redundancy} (2), \textbf{Reduce} {Duplication} (1) / {Local Complexity} (1),\\ \textbf{Simplify Parameter Handling} (4), \textbf{Simplify Suppression Warnings by Reducing the Number of Types Targeted} (1),\\ \textbf{Simplify Dependency Management} (1), \textbf{Reduce Code Duplication and Enhance Flexibility} (1), \textbf{Simplify Class Structure} (2),\\ \textbf{Remove Unnecessary Checked Exception} (1), \textbf{Reduce Dependencies and Simplify State Management} (1),\\ \textbf{Eliminate Unnecessary} {Suppression} (2) / {Variables} (2) / {Restrictions} (1)\end{tabular} \\
\hline
\textbf{\begin{tabular}[c]{@{}l@{}}Exception\\and Error Handling\end{tabular}} & \begin{tabular}[c]{@{}l@{}}\textbf{Correctness} (1), \textbf{Change Thrown Exception Type} (1), \textbf{Assert Exception Handling} (1), \textbf{Improve Error Handling} (4),\\ \textbf{Clarify Exception Semantics} (1), \textbf{Ensure Correctness of Method Overrides} (2), \textbf{Remove Unnecessary Checked Exception} (1),\\ \textbf{Ensure Correctness of Exception Handling} (1), \textbf{Improve Exception Handling Specificity} (1)\end{tabular} \\
\hline
\textbf{\begin{tabular}[c]{@{}l@{}}Type\\and Parameter Handling\end{tabular}} & \begin{tabular}[c]{@{}l@{}}\textbf{Type Safety Enhancement} (7), \textbf{Change Parameter Type} (2), \textbf{Clarify} {Parameter Semantics} (1) / {Exception Semantics} (1),\\ \textbf{Improve type specificity} (1), \textbf{Enhance Parameter Clarity and Specificity} (1)\end{tabular} \\
\hline
\textbf{\begin{tabular}[c]{@{}l@{}}Structural\\Reorganization\end{tabular}} & \begin{tabular}[c]{@{}l@{}}\textbf{Move} {Class} (1) / {Feature Module} (1), \textbf{Split Package} (1), \textbf{Rename and Reclassification} (1), {Consolidate} (1)\\ \textbf{Package Structure} {Optimization} (1) / {Consolidation} (1), \textbf{Move And Rename Attribute} (1), \textbf{Extract Interface} (1), \textbf{Simplify Hierarchy} (1),\\ \textbf{Improve Specificity in Class Responsibilities} (1)\end{tabular} \\
\hline
\textbf{\begin{tabular}[c]{@{}l@{}}Performance\\and Resource Management\end{tabular}} & \begin{tabular}[c]{@{}l@{}}\textbf{Performance Optimization} (1), \textbf{Facilitate Thread Management} (1), \textbf{Enhance} {Memory Management} (2) / {Logging and Integration} (1),\\ \textbf{Resource Management Optimization} (2), \textbf{Performance Enhancement} (1)\end{tabular} \\
\hline
\textbf{\begin{tabular}[c]{@{}l@{}}Consistency\\and Standardization\end{tabular}} & \begin{tabular}[c]{@{}l@{}}\textbf{Consistency and Metric Accuracy} (1), \textbf{Annotation Update for Deprecation} (3), \textbf{Improve Consistency} (1), \textbf{Ensure Correctness} (3),\\ \textbf{Replace Custom Abstractions with Standard Methods} (1), \textbf{Address Annotation Warnings} (1)\end{tabular} \\
\hline
\textbf{\begin{tabular}[c]{@{}l@{}}Design Principles\\and Patterns\end{tabular}} & \begin{tabular}[c]{@{}l@{}}\textbf{Enforce Design Patterns} (1), \textbf{Apply SRP} (1), \textbf{Refactor for Code Compatibility} (1), \textbf{Enhance Interface Abstraction} (1),\\ \textbf{Improve Internal Cohesion} (1)\end{tabular} \\
\hline
\textbf{\begin{tabular}[c]{@{}l@{}}Security\\and Safety\end{tabular}} & \begin{tabular}[c]{@{}l@{}}\textbf{Ensure} {Null Safety} (1) / {Thread Safety} (1), \textbf{Security and Consistency} (1), \textbf{Code Safety and Reliability} (9), \textbf{Change Method Access Level} (2),\\ \textbf{Improve Type Safety} (1)\end{tabular} \\
\hline
\textbf{\begin{tabular}[c]{@{}l@{}}Support\\(New) Functionalities\end{tabular}} & \begin{tabular}[c]{@{}l@{}}{Enhance} (3) / {Consolidate} (2) \textbf{Functionality}, \textbf{Enhance Invocation Logic} (1), \textbf{Enhance Robustness} (3), \textbf{Support New Deature} (1),\\ \textbf{Enhance Access Control} (1), \textbf{Introduce Necessary Structures for New Functionality} (1)\end{tabular} \\
\hline
\textbf{\begin{tabular}[c]{@{}l@{}}Other\\Specialized Goals\end{tabular}} & \begin{tabular}[c]{@{}l@{}}\textbf{Instance-specific Behavior Enhancement} (1), \textbf{Serialization Compatibility} (3), \textbf{Move Method} (1), \textbf{Collapse Hierarchy} (1), \textbf{Merge Attribute} (1),\\ \textbf{Leverage Standard Library} (1), \textbf{Suppress Specific Compiler Warning} (3), \textbf{Implement Null Object Pattern} (1), \textbf{Split} {Variable} (1) / {Package} (1) /\\ {Attribute} (1), \textbf{Update Attribute Value to Accommodate New Requirements or Changes} (1), \textbf{Add Modifier Correctly} (1), \textbf{Enable Interoperability} (1),\\ \textbf{Update Annotation to Reflect a Different Attribute or Dependency} (1), \textbf{Centralize Common Functionality} (1)\end{tabular} \\
\hline
\end{tabular}%
}
\end{table}

\end{document}